\documentclass[a4paper,11pt]{article}
\pdfoutput=1 

\usepackage{jheppub} 

\usepackage[T1]{fontenc} 

\newcommand{\Eq}[1]{Eq.~\eqref{#1}}
\newcommand{\Fig}[1]{Fig.~\ref{#1}}
\newcommand{\Sec}[1]{Section~\ref{#1}}
\newcommand{\C}{$C_1^{(1)}$}
\usepackage{graphicx}
\usepackage{float}
\usepackage{amsmath}
\usepackage{caption}
\usepackage{xcolor}
\usepackage{soul}
\usepackage{subcaption}
\usepackage{fancyhdr}
\usepackage{appendix}
\title{Disentangling Jet Modification in Jet Simulations and in $Z$+Jet Data}

\author[a]{Jasmine Brewer}
\author[b]{Quinn Brodsky}
\author[c]{Krishna Rajagopal}

\affiliation[a]{Theoretical Physics Department, CERN, \\CH-1211 Genève 23, Switzerland}
\affiliation[b]{Massachusetts Institute of Technology,\\
  77 Massachusetts Avenue, Cambridge, USA}
\affiliation[c]{Center for Theoretical Physics, Massachusetts Institute of Technology,\\
77 Massachusetts Avenue, Cambridge, USA}

\emailAdd{jasmine.brewer@cern.ch}
\emailAdd{quinnb@mit.edu}
\emailAdd{krishna@mit.edu}

\preprint{CERN-TH-2021-163\\\rightline{MIT-CTP/5344}}

\abstract{
In this work, we study the impact of selection biases on jet structure and substructure observables and separate these effects from effects caused by jet quenching.
We use the angular separation $\Delta R$ of the hardest splitting in a jet as the primary example of a substructure observable.
For illustrative purposes, we first conduct a simplified Monte Carlo study in which it is possible to identify the same jet after quenching in a heavy ion collision and as it would have been if it had formed in vacuum.
If we select a sample of jets by placing a cut on their quenched $p_T$ (as an experimentalist might do in heavy ion collisions) and, as is possible only in a Monte Carlo study, compare to the same jets unquenched, the $\Delta R$ distribution seems to be unmodified by quenching. However, if we select a sample of jets formed in vacuum by placing a cut on their unquenched $p_T$ and compare to those same jets after quenching, we see a significant enhancement in the number of  jets with large $\Delta R$, primarily due to the soft particles reconstructed as a part of the jet that originate from the wake in the droplet of quark-gluon plasma excited by the parton shower. We confirm that the jets contributing to this enhancement are those jets which lose the most energy, which are not included in the sample selected after quenching;
jets selected after quenching
are those which lose a small fraction of their energy. 
Next, we repeat this Monte Carlo analysis using a method that {\it is} available to experimentalists: in a sample of jets with a recoiling $Z$ boson, we show that selecting jets based on the jet $p_T$ after quenching yields a 
$\Delta R$ distribution that appears unmodified while selecting a sample of jets produced in association with a $Z$ boson whose (unmodified) $p_T$ is above some cut 
yields a significant enhancement in the number of jets with large $\Delta R$. 
We again confirm that this is due to particles from the wake, and that the jets contributing to this enhancement are those which have lost a significant fraction of their energy. Lastly, we discuss how grooming can be used to vary the importance of the contribution of medium response to the jets.}

\begin{document} 
\maketitle
\flushbottom

\section{Introduction}
\label{sec:intro}

High-energy collisions between large nuclei at the Large Hadron Collider and the Relativistic Heavy Ion Collider produce droplets of the deconfined phase of QCD matter, known as quark--gluon plasma.
In these collisions, high momentum-transfer interactions between partons in the nuclei can also produce collimated sprays of particles called jets.
Because the jets produced in a heavy ion collision plow through a droplet of quark-gluon plasma, the number of jets above a given energy can be suppressed, and the characteristics of the surviving jets can be modified, relative to what is seen in proton--proton collisions.
The discovery of parton energy loss \cite{STAR:2005gfr,PHENIX:2004vcz} and the observed suppression of jets in heavy ion collisions compared to proton--proton collisions \cite{Aad:2010bu,Chatrchyan:2011sx,Adam:2015ewa} provide crucial evidence for the formation of quark--gluon plasma.
The modification of jets in heavy ion collisions is an important probe of the properties and structure of quark--gluon plasma. 
(See Refs.~\cite{Casalderrey-Solana:2007knd,dEnterria:2009xfs,Wiedemann:2009sh,Majumder:2010qh,Mehtar-Tani:2013pia,Connors:2017ptx,Busza:2018rrf} for reviews.)

The modification of the structure and substructure of jets in heavy ion collisions provides a more detailed view into the interplay between jets and an expanding cooling droplet of quark--gluon plasma.
Jets that develop while immersed in a hot background medium have modified structure relative to those formed in vacuum, for example due to medium-induced emission~\cite{Baier:1996kr,Zakharov:1997uu,Gyulassy:2000er,Armesto:2003jh} or drag~\cite{Gubser:2006bz,Casalderrey-Solana:2006fio,Herzog:2006gh,Liu:2006ug,Chesler:2008uy,Casalderrey-Solana:2011dxg}. 
In addition, partons in the jet excite a ``wake'' in the medium they pass through~\cite{Casalderrey-Solana:2004fdk,Neufeld:2008fi,Betz:2008ka,Casalderrey-Solana:2016jvj,Tachibana:2017syd,JETSCAPE:2020uew,Casalderrey-Solana:2020rsj}. 
This wake carries momentum in the jet direction ``lost'' by the jet partons and after hadronization therefore yields  lower-$p_T$ particles from the medium that are correlated with the jet direction. As a consequence, even after subtracting backgrounds that are uncorrelated with the jet axis, jets reconstructed in heavy ion collisions include some particles originating from the wake.
Substantial recent work has focused on developing and employing novel jet observables~\cite{Dreyer:2018nbf,Andrews:2018jcm}, including to distinguish the modification of the hard structure of jets~\cite{Mehtar-Tani:2016aco,Casalderrey-Solana:2019ubu, Mehtar-Tani:2019rrk,Apolinario:2020uvt} and the contribution of medium response to observed features of jets~\cite{Casalderrey-Solana:2016jvj,Milhano:2017nzm,CMS:2021vui}.
A particularly interesting class of such observables involve using jet grooming \cite{Larkoski:2014wba} to systematically remove soft and wide-angle radiation from the jet.
In proton--proton collisions, grooming improves the perturbative calculability of jet observables~\cite{Dasgupta:2013ihk,Larkoski:2014wba}. Measurements on groomed jets can access the momentum sharing, $z_g$, and opening angle, $\Delta R$, of the first hard splitting in the shower, respectively \cite{Larkoski:2015lea, ATLAS:2019mgf,ATLAS:2018zhf,CMS:2018ypj}.
Speaking loosely, $\Delta R$ is the angular separation between the two most separated 
substructures in a jet that pass the soft drop condition (see \Sec{sec:methods} for details)
and $z_g$ is the momentum fraction carried by the softer of the two substructures.
In heavy ion collisions, due to backgrounds~\cite{Mulligan:2020tim}, the lack of angular ordering of the shower~\cite{Mehtar-Tani:2010ebp}, parton energy loss, and the response of the medium to the jet, these groomed observables are not expected to provide direct access to the first hard splitting in the shower.
However, they remain precisely defined, and powerful, observables for characterizing the substructure of quenched jets and can therefore be used to quantify how jet substructure is modified by quenching~\cite{Mehtar-Tani:2016aco,Chien:2016led, Chang:2017gkt, Casalderrey-Solana:2019ubu, Chien:2018dfn,Caucal:2019uvr}. For this reason they have been the subject of extensive recent experimental study in heavy ion collisions \cite{CMS:2017qlm,CMS:2018fof,ALICE:2021obz,ALICE:2019ykw,Oh:2020yyn}.

A natural first thought is that if we wish to see how jets are modified via passage through a droplet of quark-gluon plasma we should start by comparing the characteristics of ensembles of jets selected with the same cuts in heavy ion collisions and proton--proton collisions with the same collision energy per nucleon.  This approach does not work in a straightforward fashion, however, because
selection biases in heavy ion collisions play an important role in the phenomenology of jet modification.
The production of jets falls very steeply with energy, so selecting jets in heavy ion collisions based on their $p_T$ after energy loss in the plasma biases toward those jets that lost as little energy as possible \cite{Renk:2012ve, dEnterria:2009xfs,Milhano:2015mng,Casalderrey-Solana:2016jvj}. 
That is, if we select an ensemble of jets with $p_T$ above some cut $p_T^{\rm cut}$ in a sample of heavy ion collision events, it is  unlikely that our ensemble includes jets that lost a lot of energy since they would have had $p_T \gg p_T^{\rm cut}$ had they not been quenched, and such jets are much more rarely produced.
This selection bias that favors jets which hardly lose any energy has a large impact on the apparent modification of jet structure and substructure observables~\cite{Spousta:2015fca, Milhano:2015mng,Casalderrey-Solana:2016jvj,Rajagopal:2016uip,Brewer:2017fqy,Brewer:2018mpk,Casalderrey-Solana:2018wrw,Casalderrey-Solana:2019ubu,Caucal:2020xad}, biasing towards selecting those jets that are less modified. 
Selection biases can be reduced, but not eliminated, in dijet events~\cite{Brewer:2018dfs, Takacs:2021bpv} or using machine learning techniques~\cite{Du:2020pmp,Du:2021pqa}.

Despite their relative rarity, boson-tagged jet events are an important option for better disentangling the effects of selection bias on observables from the effects of jet modification.
In this special class of events, we can identify an ensemble of jets produced in association with an uncolored boson (photon or $Z$) whose $p_T$ is above some cut. And, since quark-gluon plasma is transparent to the boson this selection does not bias the ensemble of jets toward those that have lost less or more energy.
Loosely speaking, we can think of the $p_T$ of the boson, unmodified by the plasma, as giving us information about the energy that the recoiling jet would have had if it too were unmodified by the plasma. This picture is oversimplified, but we will show that it is also not crucial: selecting jets based upon the $p_T$ of the boson nonetheless avoids the selection bias that we are trying to disentangle ourselves from.
Boson-tagged jet measurements \cite{CMS:2017ehl, CMS:2017eqd, CMS:2018jco, ATLAS:2018dgb, ATLAS:2019dsv} thus provide important complementary information to inclusive jet measurements or measurements of dijet events.

In this paper we systematically study the effects of selection bias.
As an illustrative simplifying case for a Monte Carlo study, we first consider a sample of jets generated in \textsc{Pythia8}~\cite{Sjostrand:2007gs}
without including any effects of nuclear parton distribution functions (nPDFs); in the hybrid model of jet quenching this allows us to produce ``unquenched'' jets (as in $pp$ collisions) and subsequently compute their modification by the plasma, jet by jet, as we will discuss extensively in the next Section. We call this a ``matched jet'' sample, since for each individual unquenched jet we can look at the corresponding quenched jet {\it and vice versa}.
In this Monte Carlo analysis, we can 
compare each quenched jet with the unquenched jet that it would have been in the absence of medium effects.
With this collection of matched jets in hand, we can then select an ensemble of jets based on the $p_T$ of either the quenched or the unquenched jet.
The difference between the distribution of some jet observable in these two ensembles quantifies the role of selection bias.
It should be apparent that this analysis can only be done in a Monte Carlo study, not in an analysis of experimental data: experimentally, ensembles of jets in proton-proton and heavy ion collisions are selected based only upon their $p_T$ and are compared to each other at the distribution level. It is impossible, even in principle, to imagine accessing the properties of the same jet with and without the effects of quenching.
Nevertheless, the results of this Monte Carlo study prove to be illustrative indeed.

Next, we turn to an analysis that may be experimentally realizable by considering a sample of jets produced in association with (e.g.~recoiling against) a $Z$ boson\footnote{Although the points that we make in this paper could be made for jets produced in association with a photon originating from the same initial hard scattering that produces the jets, we focus on $Z$+jet events since all $Z$ bosons originate in that way whereas photons can originate from parton showers or from the subsequent decay of high-$p_T$ hadrons.}. We investigate the role of selection bias by comparing ensembles of jets based either upon requiring the $Z$ boson $p_T$ to be above some cut or upon requiring the (quenched) jet $p_T$ to be above some cut.

This paper is structured as follows.
In \Sec{sec:methods}, we discuss the Monte Carlo samples we use and describe the matching procedure we employ to match quenched and unquenched versions of the same jet.
In \Sec{sec:results}, we show that selecting jets based on the quenched $p_T$ substantially reduces the fractional energy loss of jets in the sample compared to selecting on the unquenched $p_T$ or the $p_T$ of the $Z$ boson.
We discuss the implications of this effect for jet structure and substructure observables, with the $\Delta R$ of groomed jets and the \C\ ~\cite{Larkoski:2013eya} (related to the angular width) of ungroomed jets as examples.
When selection biases favoring jets that lose little energy are eliminated, for example by choosing a sample of events based upon the $p_T$ of a $Z$ boson in them, we find 
that the distribution of these jet observables is substantially modified by quenching. Substantial contributions come from hadrons originating from the wake of the jets in the quark-gluon plasma that are reconstructed (as in an experimental analysis) as part of the quenched jets. In contrast,  evidence for these substantial modifications to the structure of jets introduced by quenching is
largely absent in ensembles of jets selected with a bias favoring those jets that lose the least energy, as for jets in heavy ion collisions selected based upon their $p_T$. 
In \Sec{sec:grooming}, we study how different grooming settings can be used to enhance or reduce the effects of medium response in $Z$+jet events.

\section{Monte Carlo Samples and Matching Quenched and Unquenched Jets}
\label{sec:methods}

The results in this work are based on the hybrid model of jet quenching introduced in Refs.~\cite{Casalderrey-Solana:2014bpa,Casalderrey-Solana:2015vaa,Casalderrey-Solana:2016jvj,Casalderrey-Solana:2018wrw,Casalderrey-Solana:2019ubu,Casalderrey-Solana:2020jbx}. Jets in the hybrid model are produced in \textsc{Pythia8} \cite{Sjostrand:2007gs} and the vacuum parton shower appropriate for a proton--proton collision obtained from \textsc{Pythia} can then be embedded in an evolving, boost-invariant hydrodynamic medium in which each parton in the shower loses energy. We shall use an inclusive jet sample and a $Z$+jet sample, both produced in pp and PbPb collisions with $\sqrt{s_{NN}}=5.02$~TeV. We choose a hydrodynamic medium corresponding to the $0-5\%$ most central heavy ion collisions at that $\sqrt{s_{NN}}$. 
Events in the inclusive jet sample come from an initial hard parton scattering with 
$p_T>50$~GeV, while the $Z$+jet sample was 
generated for hard processes with $p_T>5$~GeV 
with a $p_T^{-5}$ weighting to over-sample high-$p_T$ processes. 
We reconstruct jets using \textsc{Fastjet} \cite{Cacciari:2011ma} with the anti-$k_T$ algorithm \cite{Cacciari:2008gp} with a jet radius parameter of $R=0.4$. The inclusive jet sample is generated without nuclear PDFs (which enables the Monte Carlo matching procedure discussed above and, in more detail, below) and we require jets to have $|\eta|<1.8$. For the $Z$+jet sample we include nuclear PDFs and require $|\eta_\text{Z}|<2.4$. 

In the hybrid model, the final-state energy loss of partons in the shower has a dependence on path-length that is inspired by a calculation for energy loss at strong coupling in holography \cite{Chesler:2014jva,Chesler:2015nqz}. The hybrid model contains a single free parameter $\kappa_\text{sc}$ controlling the energy loss rate $dE/dx$ of a parton in the strongly-coupled plasma. The samples discussed here were generated with $\kappa_\text{sc} = 0.404$, which is the best-fit value for describing hadron and jet suppression at $T=145$~MeV without resolution effects included~\cite{Casalderrey-Solana:2018wrw}. 
We include hadronization effects and the wake that the jet leaves behind in the medium through which it propagates. However, we can distinguish between particles coming from the \textsc{Pythia} shower and particles produced from medium response, and hence can either include or neglect the effect of medium response in the jet reconstruction. The hybrid model does not include a genuine fluctuating background, meaning that background subtraction in the model~\cite{Casalderrey-Solana:2016jvj,Casalderrey-Solana:2019ubu} is not a complete representation of background subtraction as done by experimentalists; note that the latter can also have an important effect on jet structure and substructure \cite{KunnawalkamElayavalli:2017hxo,Mulligan:2020tim}.

For our illustrative, but simplified, Monte Carlo study, we consider an inclusive jet sample generated without nPDFs. In this case, we can produce ``unquenched'' jets (as in $pp$) and subsequently compute their modification by the plasma in the hybrid model. We start with a sample including every unquenched jet 
from the sample of jets from \textsc{Pythia8} described above whose reconstructed transverse momentum has
$p_T>30$~GeV. 
We then match each unquenched jet in an event to the quenched jet in the same event that has $p_T>30$~GeV that is closest in the $(\eta,\phi)$ plane, requiring that $\sqrt{\Delta \eta ^2 + \Delta \phi^2}\leq 0.4$ between the quenched and unquenched jet.
For emphasis, we restate that this is a Monte Carlo study which relies upon a matching procedure that cannot be realized, even in principle, with experimental data and also relies upon neglecting nuclear PDFs. It is additionally important to note that the \textsc{Pythia} shower in the hybrid model is not dynamically interfaced with the medium evolution, so the structure of the shower is (assumed to be) identical in the vacuum and heavy-ion samples, despite energy loss of partons in the shower. This assumption is common to many heavy ion energy loss models, but is to be contrasted with \textsc{Jewel}~\cite{Zapp:2012ak}.

\begin{figure}[t]
\centering
  \includegraphics[scale=.6]{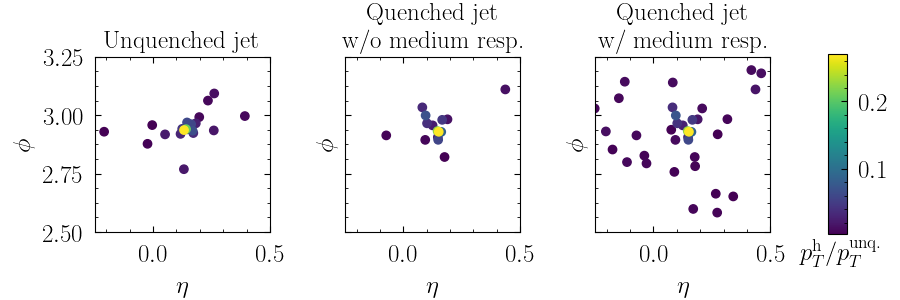}  
  \vspace{-0.06in}
\caption{Illustration exemplifying the Monte Carlo matching procedure. Left panel: $(\eta,\phi)$ distribution of hadrons in an unquenched jet, as reconstructed. Middle panel: the same jet, reconstructed after quenching but upon leaving out hadrons originating from the response of the medium, e.g. the wake the jet excites. Right panel: the same jet, reconstructed after quenching, with hadrons from the medium response included. Color indicates hadron $p_T$ as a fraction of the $p_T$ of this unquenched jet, which has $p_T^\text{unq.}=177$~GeV.
}
\vspace{-0.09in}
\label{fig:matching}
\end{figure}

In Fig.~\ref{fig:matching}, we show a visual representation of jets before and after quenching that were matched in this way. In the case of the quenched jet, we show it both with and without hadrons originating from the response of the medium to the jet. Since jets lose momentum to the medium, they leave behind a wake in the droplet of quark-gluon plasma that has a net momentum in the jet direction.  After hadronization, this wake turns into hadrons. Because their net momentum is in the jet direction, what an experimentalist reconstructs as a jet after perfect subtraction of uncorrelated backgrounds would still contain these hadrons.

We select samples of matched jets either by their quenched $p_T$ (keeping only matched jets where the quenched jet has $p_T^\text{q.}>p_T^\text{cut}$) or by their unquenched $p_T$ (keeping only matched jets where the unquenched jet has $p_T^\text{unq.}>p_T^\text{cut}$). In either case, we choose $p_T^\text{cut}=80$~GeV. We include the zero, one or two jets from each event that satisfy $p_T>p_T^\text{cut}$. 
Note that in the sample selected with $p_T^\text{unq.}>p_T^\text{cut}$, the matched quenched jets can have transverse momenta well below $p_T^\text{cut}$, as low as 30 GeV.
In cases where the quenched jet has lost so much energy that its $p_T$ has dropped below 30~GeV, the unquenched jet has no match and is removed from the sample.

Next, we propose a comparison that {\it is} realizable using experimental data. We shall see that although it is not quite as crisp as the one above, it yields similar conclusions. We consider a $Z$+jet sample generated as described above, this time with nPDFs included. 
We study selection bias by comparing results obtained from two samples. First, we select events in which $p_T^\text{Z}>p_T^\text{cut}$ where we choose $p_T^\text{cut}=80$~GeV,
and include the highest-$p_T$ jet in each such event in our sample. If no jet with $p_T>30$~GeV is reconstructed in an event we discard that event from the analysis. The jet $p_T$ may be much lower than $p_T^\text{cut}$, although it will have  $p_T>30$~GeV.
For our second sample, we select jets in events with a $Z$ boson
where the jet satisfies $p_T^\text{jet}>p_T^\text{cut}$, as before choosing $p_T^\text{cut}=80$~GeV.
Although they are rare in this sample, by analogy with the procedure that we followed in our inclusive jet analysis we exclude any events in which the $Z$ boson transverse momentum is below 30~GeV.

The soft drop algorithm~\cite{Larkoski:2014wba} that we employ to construct the groomed observables that we analyze in all these samples has two parameters $z_\text{cut}$ and $\beta$ that determine the extent of the grooming. 
After clustering jets with any algorithm (here, anti-$k_t$), constituents are iteratively reclustered with those closest in the $(\eta,\phi$) plane using the Cambridge-Aachen algorithm.
The soft drop algorithm then goes through the reclustered tree recursively, discarding the softer branch until
\begin{equation}
\label{eq:softdrop}
\frac{\min(p_{T,1},p_{T,2})}{p_{T,1}+p_{T,2}} > z_{\text{cut}}\left(\frac{\Delta R_{12}}{R}\right)^\beta\,.
\end{equation}
Here, $p_{T,1}$ and $p_{T,2}$ are the transverse momenta of subjets 1 and 2 and $\Delta R_{12}$ is their angular separation in the $(\eta,\phi)$ plane.
$\Delta R$ and $z_g$ refer to the angular separation and the momentum sharing of the earliest splitting that satisfies the condition \Eq{eq:softdrop}.
Except in \Sec{sec:grooming}, we will show results for jets groomed with $z_\text{cut}=0.1$ and $\beta=0$.
In \Sec{sec:grooming} we will further explore the influence of different grooming settings on the contribution of particles originating from medium response.

\section{Role of Selection Bias in Jet Observables}
\label{sec:results}

In this Section, we study the role of selection bias in jet substructure observables, using the $\Delta R$ distribution of jets groomed with $z_\text{cut}=0.1$, $\beta=0$, and the \C distribution of ungroomed jets as illustrative examples. In different ways, both of these observables are measures of the angular width of a jet.
In vacuum, the $\Delta R$ of a jet has the interpretation as the opening angle of the first, largest angle, hard splitting. 
More generally, it is the angular separation between the two most separated substructures in a jet, as identified via the soft drop procedure~\cite{Larkoski:2014wba}.
\C is defined from the angular distribution of energy in an ungroomed jet and is more directly related to the jet width~\cite{Larkoski:2013eya}.

In all subsequent Figures, we shall show results for jets selected based on their quenched $p_T$ (both for the matched jets from our inclusive jet sample, and for the jets in our  $Z$+jet sample) in blue. And, we shall show results for jets selected based on their unquenched $p_T$ (for matched jets) or the $p_T$ of the recoiling $Z$ boson (for $Z$+jet) in orange.
As we discussed in the Introduction, 
the $p_T$ of the $Z$ boson does not tell us the unquenched jet $p_T$ in any straightforward way.
We use the same color to depict results for matched jets selected based upon their unquenched $p_T$ and $Z$+jet results for jets selected based upon the $p_T$ of the $Z$ only because both these selection methods are free from any bias favoring jets that lose less energy. We shall see that eliminating this selection bias has similar consequences in the two cases.
As described in the previous Section, we impose a selection cut of $80$~GeV on the appropriate $p_T$, and additionally require all jets considered to have $p_T>30$~GeV.
Dotted lines in all subsequent Figures show the distributions of observables for unquenched jets in vacuum,  dashed lines show the distributions for jets quenched by the medium produced in a heavy ion collision but without including the hadrons coming from the wake in the medium induced by the jet, and solid lines show the distributions for quenched jets in a heavy ion collision including the hadrons coming from the response of the medium. 

\begin{figure}[t]
\begin{subfigure}{.45\textwidth}
  \includegraphics[scale=.38]{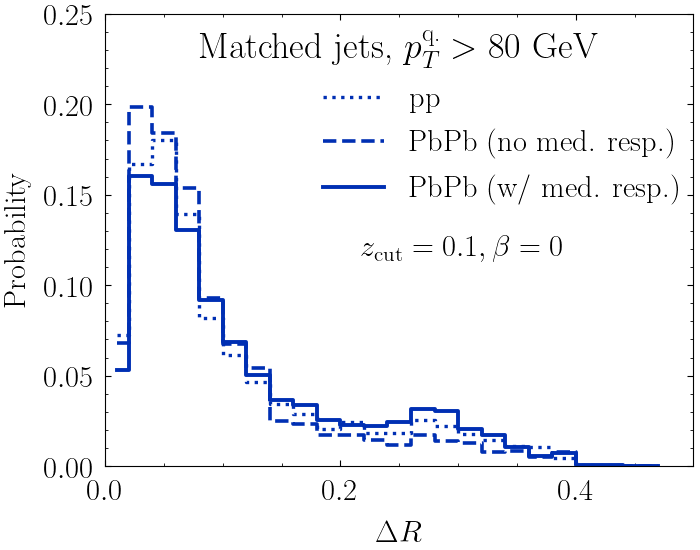}  
  \caption{}
  \label{fig:fig2a}
\end{subfigure}
\begin{subfigure}{.45\textwidth}
  \includegraphics[scale=.38]{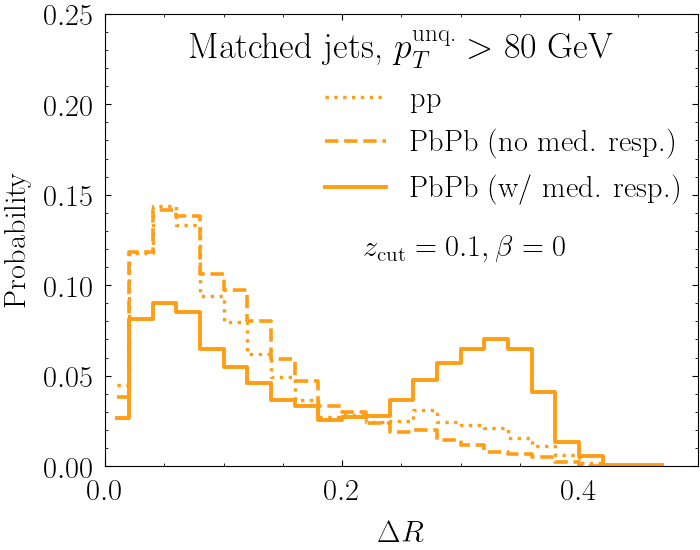}  
  \caption{}
  \label{fig:fig2b}
\end{subfigure}
\centering
\vspace{-0.06in}
\caption{(a) $\Delta R$ distributions for a sample of jets selected on the basis of having quenched $p_T$ above $80$~GeV (solid and dashed blue) 
and the matching unquenched jets (dotted blue).  Each quenched jet was matched to the unquenched jet that it would have been if it had formed in vacuum. 
(b) $\Delta R$ distributions for a sample of jets selected on the basis of having unquenched $p_T$ above $80$~GeV (dotted orange) and the matching quenched jets (solid and dashed orange).
Dashed curves are for jets quenched as in a heavy ion collision with hadrons originating from medium response artificially excluded; solid curves are for quenched jets including those hadrons from medium response that are reconstructed as a part of the jets.
}
\vspace{-0.09in}
\label{fig:fig2}
\end{figure}

In \Fig{fig:fig2}, we show the distributions of $\Delta R$ for samples of matched jets selected in these two different ways, blue and orange. We emphasize that the selection biases are identical for all the blue curves, and are identical for all the orange curves, since in either the blue samples or the orange samples every selected quenched jet is matched with an unquenched jet or vice versa.  The selection biases are very different for blue relative to orange since the blue samples were selected on the basis of the quenched jet having $p_T>p_T^{\text{cut}}=80$~GeV, which biases the blue sample 
strongly in favor of those jets that lose the least energy.
This kind of selection bias is sometimes called a ``survivor bias'', since the sample includes only those jets that ``survive quenching'', where here ``survive'' means keep their transverse momentum above $p_T^{\text{cut}}$. Jets that lose more energy are not selected.  This bias is absent in the orange samples, where the sample of quenched jets includes jets whose unquenched $p_T$ was above the cut but which have lost a lot of $p_T$ due to quenching.

For both selections, blue and orange, the $pp$ (dotted) and PbPb without medium response (dashed) $\Delta R$ distributions agree remarkably well.
Seeing this in the blue curves alone would leave us unsure whether the message is that the selection bias inherent in the blue samples could be responsible (meaning that this lack of modification in the $\Delta R$ distribution is only characteristic of a sample biased towards jets that lose the least energy) or whether the message could be that in the hybrid model jet quenching results in little modification of the $\Delta R$ of a jet, if particles originating from medium response are artificially excluded from the jet.
Because we see that the orange dashed curve also agrees 
so well with the orange dotted curve, we can now reach the latter conclusion with confidence since there is no bias toward jets that lose less energy in the orange samples.
One effect of the selection bias in the blue samples can be seen from the fact that all distributions in \Fig{fig:fig2a} are narrower than those in \Fig{fig:fig2b}: by selecting on the quenched $p_T$ as in \Fig{fig:fig2b}, one selects those (typically narrower) jets that are less modified, compared to selecting on the unquenched $p_T$ as in \Fig{fig:fig2b}.
We learn that it is the narrower jets that lose less energy; the ``survivor bias'' turns out to also be a bias in favor of selecting narrower jets with smaller $\Delta R$.

The story becomes more dramatic for the solid curves in \Fig{fig:fig2}, when the hadrons originating from the wake in the droplet of quark-gluon plasma excited by the passing jet are included (as must be the case for jets as reconstructed from experimental data).
There is a substantial contribution of jets with large $\Delta R$ in the solid orange curve in \Fig{fig:fig2b} that is not seen in the solid blue curve in \Fig{fig:fig2a} (which is quite similar to the blue dashed and blue dotted curves).
This contribution also makes the solid orange curve very different from the dotted orange curve.
When we include hadrons from medium response, as is the case in jets reconstructed from experimental data, our conclusion becomes: (i) in the hybrid model, jet quenching {\it does} result in a substantial modification of the $\Delta R$ of a jet, compare orange solid to orange dotted; but (ii) this effect is essentially made invisible by the selection bias inherent in selecting jets whose quenched $p_T$ is above some cut, as in the blue curves. Conclusion (i) provides a dramatic illustration of the fact that in heavy ion collisions $\Delta R$ does not describe the first hard splitting during the formation of the jet in any straightforward fashion.
Conclusion (ii) indicates that the jets in the solid orange distribution whose $\Delta R$ has been substantially
increased by quenching, and the consequent medium response, are at the same time jets that have lost a lot of energy and hence were not included in the blue sample.

\begin{figure}[t]
\centering
\begin{subfigure}{\textwidth}
\centering
\includegraphics[scale=0.38]{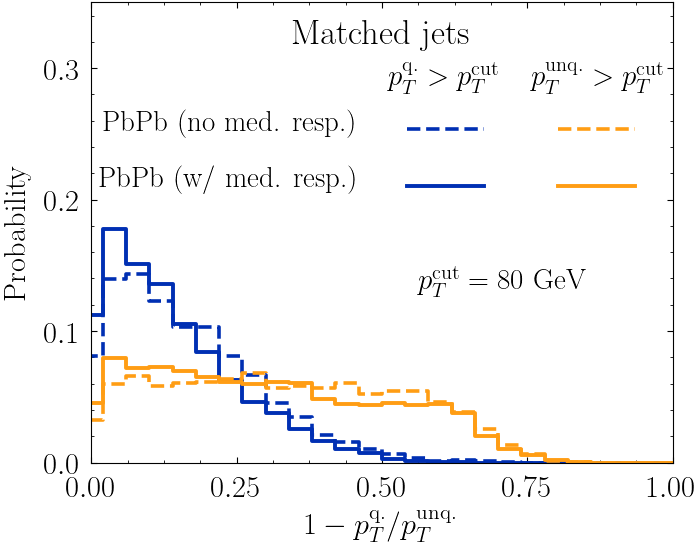}
\caption{}
\label{fig:fig3a}
\end{subfigure}\\[2ex]
\begin{subfigure}{.45\textwidth}
\centering
\includegraphics[scale=0.38]{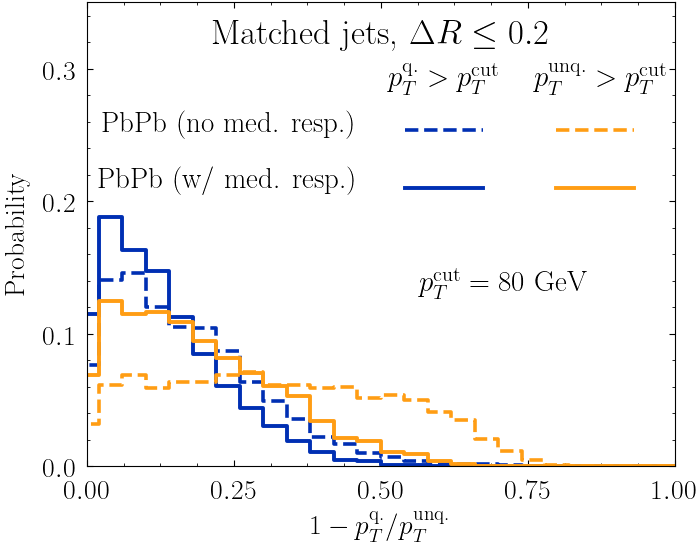}
\caption{}
\label{fig:fig3b}
\end{subfigure}%
\begin{subfigure}{.45\textwidth}
\centering
\includegraphics[scale=0.38]{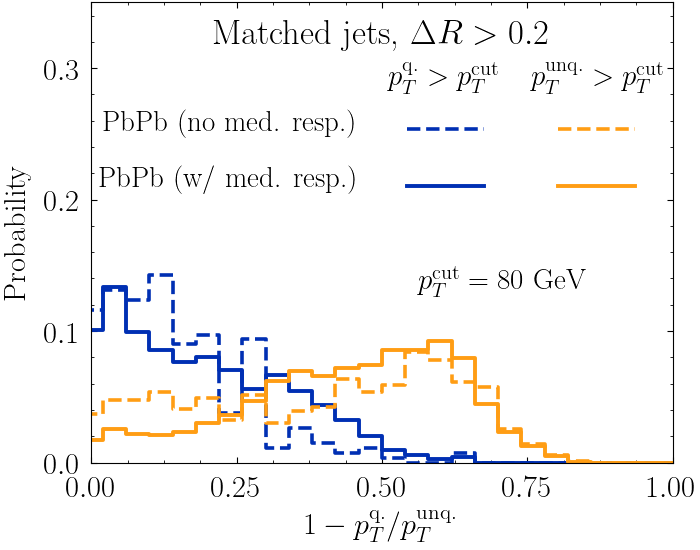}
\caption{}
\label{fig:fig3c}
\end{subfigure}
\caption{
(a) Fractional energy loss for matched jets with quenched $p_T$ above $80$~GeV (blue) and unquenched $p_T$ above $80$~GeV (orange) in heavy ion collisions without (dashed) and with (solid) medium response. Blue and orange samples are as in \Fig{fig:fig2}, as is the meaning of dashed versus solid. Panels (b) and (c) show the same distributions for those jets with $\Delta R \leq 0.2$ and $\Delta R > 0.2$, respectively.}
\label{fig:fig3}
\end{figure}

We have argued that the contribution at large $\Delta R$ seen in the solid orange curve in  \Fig{fig:fig2b} comes from those jets that have lost a large fraction of their energy and are therefore not represented in \Fig{fig:fig2a} due to the selection bias inherent in the blue samples.
This claim is substantiated in \Fig{fig:fig3}, as we explain below.
We show in \Fig{fig:fig3a} the distributions of the $p_T$ asymmetry between a quenched jet and its matched unquenched jet
(that is, the fractional energy loss experienced by the jet due to quenching) for jets selected based on their quenched $p_T$ (blue) and unquenched $p_T$ (orange).
The selection bias introduced by selecting on the quenched $p_T$ can be seen by the fact that the blue distributions in \Fig{fig:fig3a} are much closer to zero than the orange distributions.
This confirms that the bias in the blue samples is indeed a bias toward jets that lose less energy.
\Fig{fig:fig3b} and \Fig{fig:fig3c} show the fractional energy loss distributions for jets in the blue and orange samples with $\Delta R \leq 0.2$ and $\Delta R > 0.2$, respectively, inspired by isolating the prominent enhancement at large $\Delta R$ in the solid orange curves in \Fig{fig:fig2b}.
When including medium response, the jets with large $\Delta R$ (contributing to the enhancement in \Fig{fig:fig2b}) also have much larger average energy loss than those with small $\Delta R$.
This is seen by the solid orange distributions being peaked at larger fractional energy loss in \Fig{fig:fig3c} than in \Fig{fig:fig3b}.  This substantiates our claim. 
It is important to keep in mind that, when medium response is not included, jets with $\Delta R>0.2$ make up only a very small fraction of all jets, unlike when medium response is included.
Therefore the dashed orange curves in \Fig{fig:fig3b} and \Fig{fig:fig3c} should be interpreted with caution; we provide them only for completeness.

We note that measurements of $\Delta R$ have been reported in Ref.~\cite{ALICE:2021obz} and show good agreement with  previous calculations in the hybrid model provided for that publication. The ratio of the solid and dotted blue curves in \Fig{fig:fig2a} are fully comparable to the hybrid model curves depicted in green in Fig.~3 of~\cite{ALICE:2021obz}, with the exception that the kinematic cuts and grooming parameters are somewhat different. We therefore hope that the conclusions 
 in the present work may aid in interpreting this measurement by highlighting the degree to which effects of selection bias contribute to what is seen.

Returning to the comparison between the solid orange and solid blue curves in \Fig{fig:fig3a}, we remark that the solid orange curve is in a sense of greater interest since it provides a fair representation of the distribution of the fractional energy loss experienced by jets due to their passage through the quark-gluon plasma produced in heavy ion collisions.
The solid blue curve shows how dramatically the fractional energy loss, which is to say the consequence of the interaction between the partons in the jets and the quark-gluon plasma,   is lessened by selecting a sample of quenched jets with $p_T$ above some cut.  
Unfortunately, any inclusive sample of jets 
in real heavy ion collision data
selected via cuts on the jet $p_T$ must include a selection bias that makes the distribution of the fractional energy
loss in the sample analogous to that in our blue curve.

\begin{figure}[t]
\begin{subfigure}{.45\textwidth}
  \includegraphics[scale=.38]{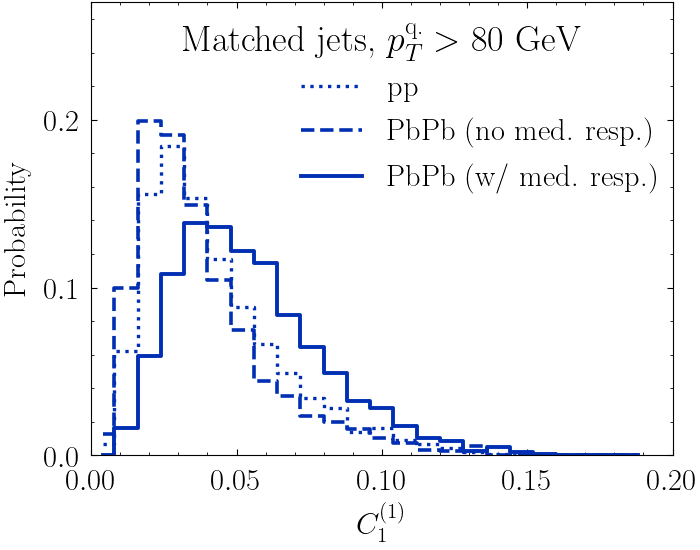}  
  \caption{}
  \label{fig:fig4a}
\end{subfigure}
\begin{subfigure}{.45\textwidth}
  \includegraphics[scale=.38]{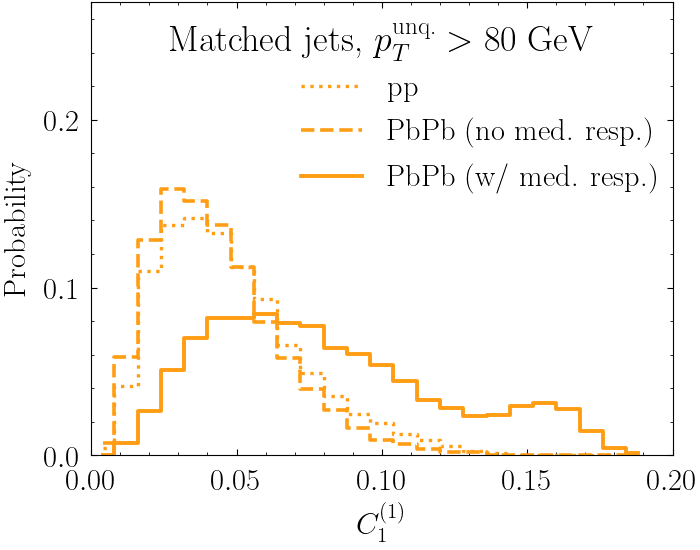}  
  \caption{}
  \label{fig:fig4b}
\end{subfigure}
\centering
\vspace{-0.06in}
\caption{\C distributions for jets with quenched $p_T$ above $80$~GeV (a) and unquenched $p_T$ above $80$~GeV (b), in vacuum (dotted), in heavy ion collisions without medium response (dashed), and in heavy ion collisions with medium response (solid).}
\vspace{-0.09in}
\label{fig:fig4}
\end{figure}

Though we have so far illustrated the consequences of selection bias, quenching, and medium response via their effects on the $\Delta R$ distribution, we wish to emphasize that our conclusions are not specific to this observable or to the grooming necessary to obtain it. 
In \Fig{fig:fig4} we repeat the analysis of \Fig{fig:fig2} using the \C observable, a measure of the angular width of {\it ungroomed} jets.
We show the \C distributions for jets selected based on the quenched $p_T$ (blue; \Fig{fig:fig4a}) and based on the unquenched $p_T$ (orange; \Fig{fig:fig4b}).
As in \Fig{fig:fig2}, there is a significant enhancement in the contribution of jets with large \C to the solid orange distribution, where we have eliminated bias favoring jets that lose less energy and included the hadrons originating from medium response as part of the reconstructed jets. 
This effect is almost eliminated by the selection biases present in the blue sample, meaning that the jets whose \C has increased in the solid orange distribution are jets that have lost a significant amount of energy.
As for $\Delta R$, the \C of jets is significantly modified by quenching, as long as we make sure to include hadrons originating from medium response and as long as we avoid biases favoring the selection of jets that have lost little energy.

\begin{figure}[t]
\begin{subfigure}{.45\textwidth}
  \includegraphics[scale=.38]{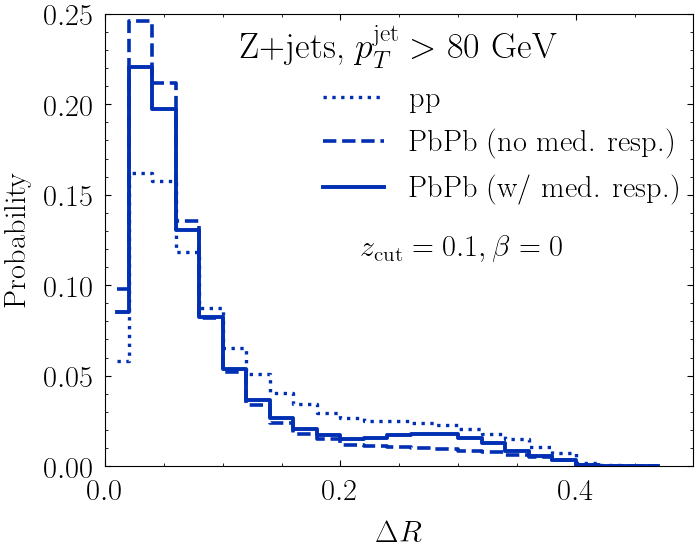}  
  \caption{}
  \label{fig:fig5a}
\end{subfigure}
\begin{subfigure}{.45\textwidth}
  \includegraphics[scale=.38]{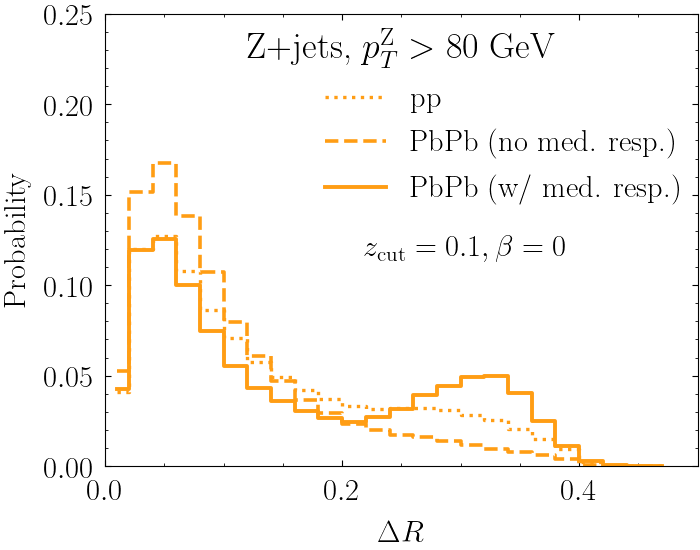}  
  \caption{}
  \label{fig:fig5b}
\end{subfigure}
\centering
\vspace{-0.06in}
\caption{$\Delta R$ distributions, (a, blue curves) for jets with $p_T$ above $80$~GeV in $Z$+jet events and (b, orange curves) for jets in events containing a $Z$ boson with $p_T$ above $80$~GeV. We show these distributions for $Z$+jet events in vacuum (dotted), in heavy ion collisions with hadrons coming from the wake that the jet deposits in the droplet of quark-gluon plasma artificially excluded (dashed), and in heavy ion collisions including  hadrons coming from the response of the medium that are reconstructed as a part of the jets (solid).}
\vspace{-0.09in}
\label{fig:fig5}
\end{figure}

Although the samples of matched jets 
selected and used in \Fig{fig:fig2} is very useful for investigating the impact of selection biases, selecting samples in this fashion is not physically realizable in any analysis of experimental data. We therefore present in \Fig{fig:fig5} a similar analysis using jets from $Z$+jet events selected in two different ways, both of which could be employed by experimentalists.
\Fig{fig:fig5} shows the $\Delta R$ distributions for samples of jets that are either selected based on the $p_T$ of the jet (\Fig{fig:fig5a}) or based on the $p_T$ of the $Z$ boson (\Fig{fig:fig5b}).
As in \Fig{fig:fig2b}, there is an enhancement at large $\Delta R$ in \Fig{fig:fig5b} originating from the response of the medium to the jets that is absent in \Fig{fig:fig5a}.
Selecting jets with $p_T$ above some cut, as in the blue curves of \Fig{fig:fig5a}, biases our sample toward jets that lose little energy. They hence excite almost no wake in the droplet of QGP and consequently yield similar $\Delta R$ distributions for $Z$+jet events in vacuum or in heavy ion collisions, whether or not
the hadrons coming from the wake are included.
In contrast, when we select $Z$ bosons with $p_T$ above some cut as in the orange curves of \Fig{fig:fig5b}, we see that the $\Delta R$ distributions of the recoiling jets {\it are} modified by quenching, in particular when the hadrons coming from the wake are included in the analysis. 
It is also noteworthy that the $\Delta R$ distributions that were remarkably similar between $pp$ and PbPb without medium response in the matched jet sample of \Fig{fig:fig2} have substantial differences in this case.
This is presumably due to the fact that jets with the same $p_T$ are not all the same: whereas the matched jet sample identifies identical jets with or without quenching, here we 
compare jets above a cut in $Z$+jet events (blue curves) to
jets in $Z$+jet events where the $Z$ is above that same cut (orange curves).

\begin{figure}[t]
\centering
\begin{subfigure}{\textwidth}
\centering
\includegraphics[scale=0.38]{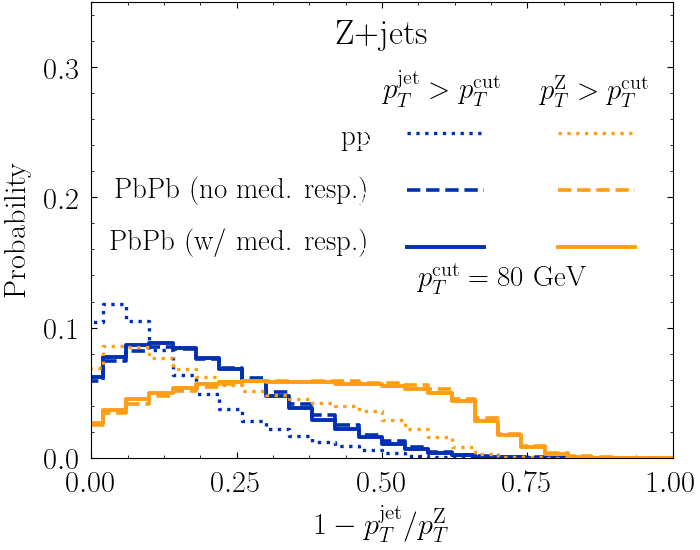}
\caption{}
\label{fig:fig6a}
\end{subfigure}\\[2ex]
\begin{subfigure}{.45\textwidth}
\centering
\includegraphics[scale=0.38]{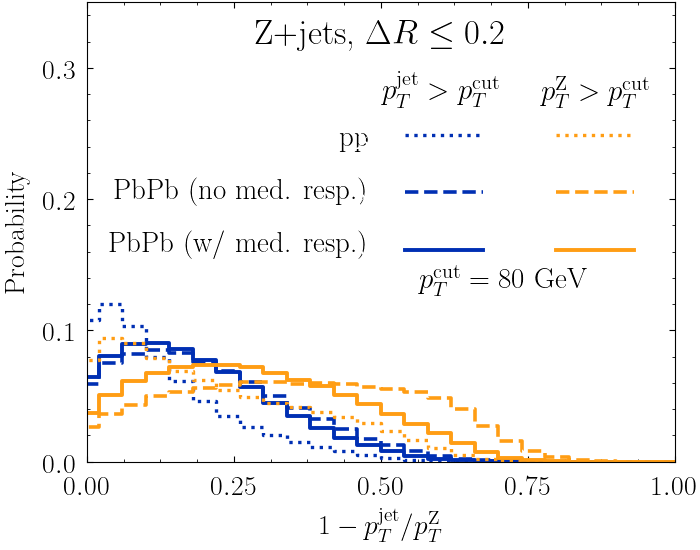}
\caption{}
\label{fig:fig6b}
\end{subfigure}%
\begin{subfigure}{.45\textwidth}
\centering
\includegraphics[scale=0.38]{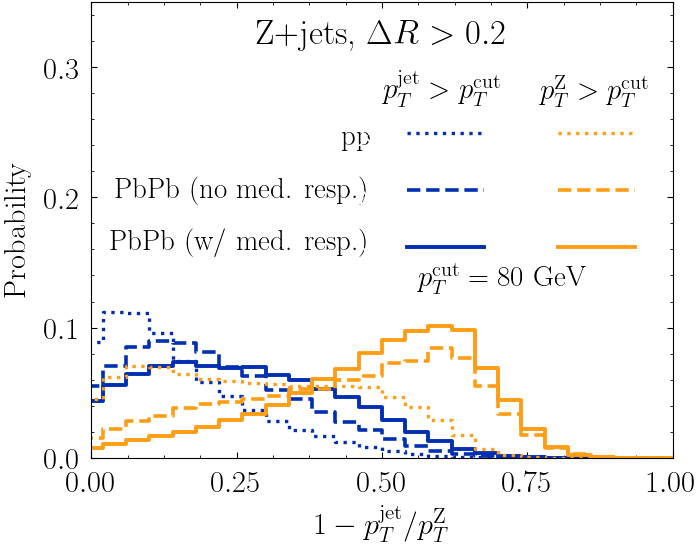}
\caption{}
\label{fig:fig6c}
\end{subfigure}
\caption{
		(a) Fractional $p_T$ asymmetry between a $Z$ boson and its recoiling jet, for jets with jet $p_T$ above $80$~GeV (blue) and $Z$ $p_T$ above $80$~GeV (orange) in vacuum (dotted), and in heavy ion collisions with hadrons originating from medium response artificially excluded from (dashed) or included in (solid) the reconstructed jets. Panels (b) and (c) show the same distributions for those jets with $\Delta R \leq 0.2$ and $\Delta R > 0.2$, respectively.}
\label{fig:fig6}
\end{figure}

In both \Fig{fig:fig2} and \Fig{fig:fig5} the selection in orange does not bias the sample towards jets that lose little energy whereas the selection in blue does. In the matched jet analysis of \Fig{fig:fig2} there is no other difference between the blue and orange samples, but that cannot be the case in \Fig{fig:fig5}. Although the analysis in \Fig{fig:fig2} is the cleanest possible way to look at the effects of selection bias, the analysis in \Fig{fig:fig5} has the great virtue of being realizable. We urge experimentalists analyzing $Z$+jet events in heavy ion collisions to perform both the blue selection and the orange selection, in pp data and in PbPb data, to see how they differ.
An enhancement in the $\Delta R$ distribution at large $\Delta R$ in the orange analysis could be an indication of the importance of medium response
to the modification of jets, and a demonstration that such modifications are obscured by selection bias in the blue analysis.

As before, we confirm in \Fig{fig:fig6} that the enhancement at large $\Delta R$ in \Fig{fig:fig5b} is from jets that lose a large fraction of their energy.
An important difference with respect to the matched jet sample, however, is that a $Z$ boson and its recoiling jet do not have the same $p_T$ even in $pp$.
Therefore we show additionally in \Fig{fig:fig6} the distribution of fractional energy difference of the $Z$ and jet in $pp$.
As before, selecting on the $Z$ boson $p_T$ (orange) rather than the jet $p_T$ (blue) substantially enhances the contribution of jets that lost a large fraction of their energy due to quenching. Also as before, jets with larger $\Delta R$ tend to experience a larger fractional energy loss than those with smaller $\Delta R$, particularly when selecting jets based on the $Z$ boson $p_T$.   The big advantage here is that samples selected as in either the solid blue curve or the solid orange curve in \Fig{fig:fig6a} {\it can} be realized in an actual analysis of $Z$+jet data from experiment, whereas an inclusive jet comparison between orange and blue samples as in our \Fig{fig:fig3a} is impossible with real data. We note again that the $Z$ boson $p_T$ and jet $p_T$ are substantially different in $pp$ collisions; the former does not tell us the latter. Regardless, we see that selecting events based upon the $Z$ boson $p_T$ in heavy ion collisions gives qualitatively similar results to selecting on the unquenched jet $p_T$ in a matched sample, because in both cases we eliminate the selection bias toward jets that lose less energy. We also note
that the interpretation of $Z$+jet event data is also more straightforward in an additional way: since the $Z$ produces no wake, complications arising from the wake of one jet 
affecting how a different jet in the same event is reconstructed~\cite{Pablos:2019ngg} are much reduced.

\begin{figure}[t]
\begin{subfigure}{.45\textwidth}
  \includegraphics[scale=.38]{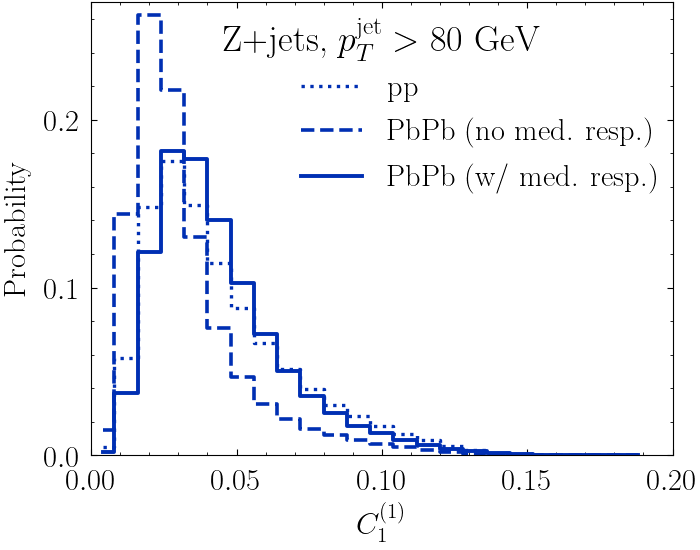}  
  \caption{}
  \label{fig:fig7a}
\end{subfigure}
\begin{subfigure}{.45\textwidth}
  \includegraphics[scale=.38]{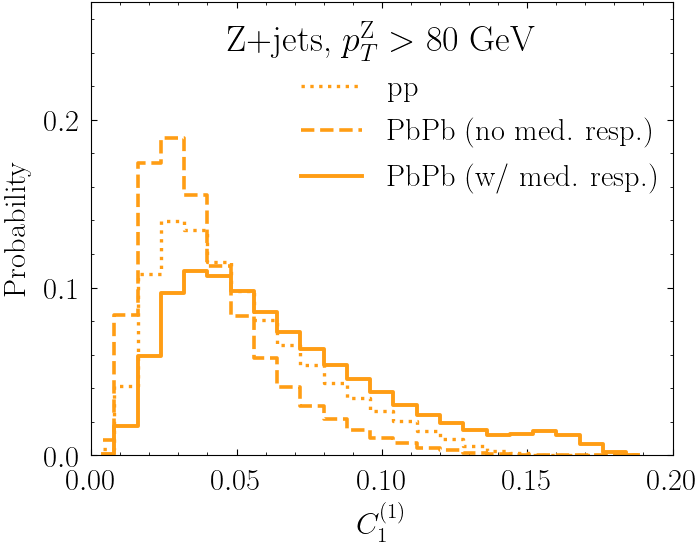}  
  \caption{}
  \label{fig:fig7b}
\end{subfigure}
\centering
\vspace{-0.06in}
\caption{
\C distributions, (a) for jets with $p_T$ above $80$~GeV and (b) for jets recoiling against a $Z$ boson with $p_T$ above $80$~GeV, in vacuum (dotted), in heavy ion collisions without medium response (dashed), and in heavy ion collisions with medium response (solid).}
\vspace{-0.09in}
\label{fig:fig7}
\end{figure}

In \Fig{fig:fig7} we show the \C distributions for jets from $Z$+jet events selected based on the jet $p_T$ (\Fig{fig:fig7a}) and based on the $Z$ boson $p_T$ (\Fig{fig:fig7b}).
The \C distributions are narrower and closer to zero in the first selection than the second, confirming that here as in \Fig{fig:fig4} the biases inherent in the way the blue sample is selected yield a sample of jets biased toward smaller \C. That is, jets with smaller \C are more likely to survive quenching with little energy loss, indicating that narrower jets lose less energy than wider jets with the same $p_T$.

\section{Impact of Grooming on Medium Response}
\label{sec:grooming}

In this Section, we explore how different grooming settings can either increase or decrease the enhancement at large $\Delta R$.
As we saw in \Sec{sec:results}, removing selection biases toward jets that lose little energy reveals a contribution to the modification of jets by quenching that comes from the response of the medium to the jet. Grooming settings that serve to groom away more of these hadrons coming from the wake should decrease the enhancement at large $\Delta R$. 
For fixed $z_\text{cut}$, increasing (decreasing) $\beta$ includes more (grooms away more) soft and wide-angle particles, while for fixed $\beta$ decreasing (increasing) $z_\text{cut}$ includes more (grooms away more) softer particles.

\begin{figure}[t]
\begin{subfigure}{.45\textwidth}
  \includegraphics[scale=.4]{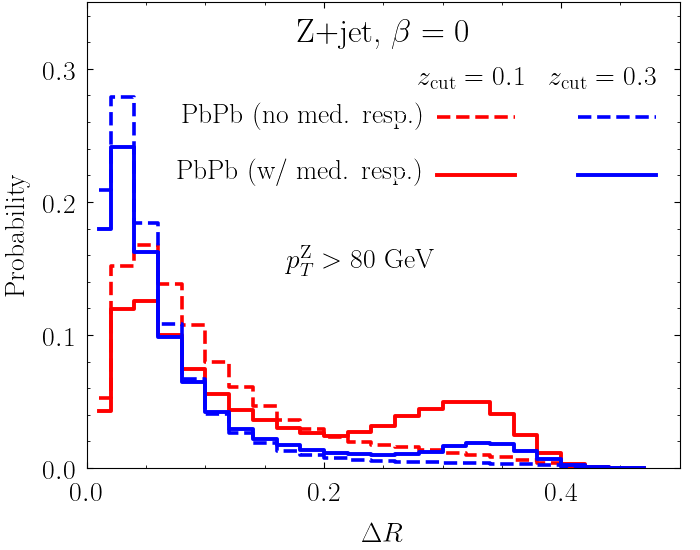}  
  \caption{}
  \label{fig:fig8a}
\end{subfigure}
\begin{subfigure}{.45\textwidth}
  \includegraphics[scale=.4]{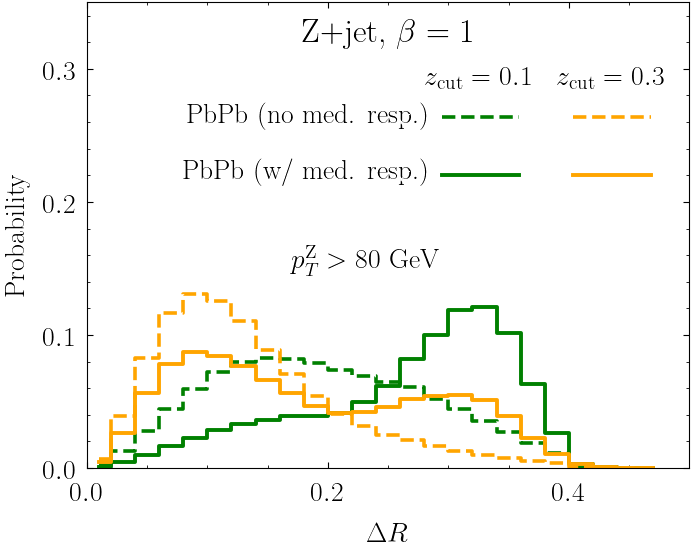}  
  \caption{}
  \label{fig:fig8b}
\end{subfigure}
\centering
\vspace{-0.06in}
\caption{$\Delta R$ distributions with and without medium response for $z_{\text{cut}}=0.1,0.3$, $\beta=0$ (a) and $z_{\text{cut}}=0.1,0.3$, $\beta=1$ (b), for jets recoiling against a $Z$ boson with $p_T$ above $80$~GeV (namely, jets selected as in the orange curves of \Fig{fig:fig5b}). All curves are for quenched jets as in PbPb collisions. Dashed curves indicate in-medium jets where hadrons originating from the wake in the medium have been artificially excluded, whereas solid curves indicate those where particles from the response of the medium are included.}
\vspace{-0.09in}
\label{fig:fig8}
\end{figure}

In \Fig{fig:fig8} we show the effect of medium response on the $\Delta R$ distributions for $Z$ boson-tagged jet samples (selected by requiring the $Z$ boson $p_T$ to be above 80 GeV, as in the orange curves in \Sec{sec:results}) with all combinations of $z_{\text{cut}}=0.1$ or $0.3$ and $\beta=0$ or $1$.
By comparing the red lines in Fig. \ref{fig:fig8a} with the green lines in Fig. \ref{fig:fig8b}, or by comparing the blue lines in Fig. \ref{fig:fig8a} with the orange lines in Fig. \ref{fig:fig8b}, we see that for a fixed $z_\text{cut}$, the effect of increasing $\beta$ corresponds to softer grooming.
That is, more soft medium response particles are included if $\beta$ is larger. Now, fix $\beta$ and let $z_\text{cut}$ vary: by comparing the red and blue lines of Fig. \ref{fig:fig8a}; or by comparing the green and orange lines of Fig. \ref{fig:fig8b}, we see that increasing $z_\text{cut}$ corresponds to grooming away more of the medium response particles, as the effect of medium response is larger in red than in blue, and larger in green than in orange.
This figure shows that one can change the $\Delta R$ distribution and the magnitude of the enhancement of large $\Delta R$ by tuning the values of $z_\text{cut}$ and $\beta$. 
Soft Drop grooming settings of $z_{\text{cut}}=0.3$ and $\beta=0$ decrease the effect of medium response by grooming away the particles that result from it, while $z_{\text{cut}}=0.1$ and $\beta=1$ keep more of these particles and hence yield larger effects of medium response in the observed $\Delta R$ distribution.

\section{Conclusion and Discussion}
\label{sec:conclusions}

In this paper we have illustrated the impact of selection biases on jet structure and substructure observables.
In a simplified Monte Carlo setup, we identified quenched and unquenched versions of the same jet.
Jets selected based on their quenched $p_T$ lose much less energy than those selected based on their unquenched $p_T$: the former selection method introduces a bias that favors those jets that lose the least energy whereas the latter selection method does not introduce this bias.
This has dramatic consequences for the $\Delta R$ and \C distributions.
The selection bias that results from selecting jets whose quenched $p_T$ is above some cut  pushes the distributions of both $\Delta R$ and \C toward smaller values (meaning that among jets with a given $p_T$, it is the narrower jets that lose the least energy).
It also substantially reduces the effect of medium response on the distributions (since jets that lose the least energy create the smallest wake in the QGP).

Next, we found qualitatively similar results in a more realistic, experimentally realizable, study of $Z$-tagged jet events. 
By selecting jets based on requiring that the $p_T$ of the recoiling $Z$ boson is above some cut we eliminate the bias favoring those jets with a given $p_T$ that lose the least energy.  
Selecting jets by cutting only on the $Z$ boson $p_T$  gives a more unbiased view of how jet quenching modifies $\Delta R$ and \C . 
When selection biases (those arising from selecting jets whose own $p_T$ is above a cut) are eliminated, the hadrons originating from the response of the medium contribute quite significantly to the 
modification of jets due to quenching, and serve to enhance the distributions at large values of $\Delta R$ and \C.
Last, we showed that different grooming settings can be used to vary the importance of medium response contributions in $Z$+jet studies.

These results have important consequences for the interpretation of jet modification observables in heavy ion collisions.
Selection biases 
that favor selecting those jets with a given $p_T$ that have lost the least energy substantially reduce the contribution from medium response to jet observables. This makes sense, since jets that lose less energy also create less of a wake.
The same selection biases also
shift the observed distributions of both $\Delta R$ and \C to lower average values, meaning that the jets that lose the least energy are those with smaller $\Delta R$ and \C, which is to say those that are narrower.
Furthermore, to the extent that jets do lose energy and excite a wake in the plasma, the contribution to $\Delta R$ from hadrons originating from this wake 
obscures the connection between the  $\Delta R$ measured in heavy ion collisions and the physics of the first hard splitting.

In this paper, we have shown results from two kinds of model studies that provide evidence for these conclusions: comparisons between matched quenched and unquenched jets from inclusive events, and comparisons between two samples of jets from $Z$+jet events, each selected in different ways.  The former approach is only possible in a Monte Carlo analysis;
it is impossible to know what a quenched jet in a heavy ion collision
would have looked like if it had been produced in vacuum. The latter approach, though, {\it can} be employed by experimentalists.  The analyses whose results we have shown in the solid blue and orange curves in Figs.~\ref{fig:fig5}, \ref{fig:fig6} and \ref{fig:fig7} as well as all four solid curves in Fig.~\ref{fig:fig8} can, and we urge should, all be done with real $Z$+jet data.  The ``orange'' analysis can be done with a $Z$+jet sample selected in a conventional way.  The ``blue'' analysis requires selecting a large sample of events containing a jet with $p_T>p_T^{\rm cut}$, for example $p_T>80$~GeV, and then keeping only the (small) fraction of those events that also contain a $Z$ boson. Although unconventional, this can certainly be done. The motivation for doing these $Z$+jet analyses is the opportunity to compare the ``orange'' and ``blue'' distributions for $\Delta R$ and \C obtained from experimental data with each other. Doing so will demonstrate the effects of the particular selection bias present in the blue analysis and absent in the orange analysis in real data, rather than in the model analyses that we have presented as motivation. If the qualitative differences point in the same direction that our results in Figs.~\ref{fig:fig5}, \ref{fig:fig6a} and \ref{fig:fig7} do, this would constitute an experimental demonstration that narrower jets lose less energy than wider jets with the same $p_T$.
Furthermore, if the shape of the orange results obtained from experimental data for the comparison between the fractional $p_T$ asymmetry at small and large $\Delta R$ are in qualitative agreement with what we have found in Figs.~\ref{fig:fig6b} and \ref{fig:fig6c}, and if the dependence of the $\Delta R$ distribution on choices of grooming parameters is in qualitative agreement with what we have found in Fig.~\ref{fig:fig8}, this would provide strong evidence from experiment that hadrons coming from medium response are central to the modification of $\Delta R$ in those jets that lose significant energy and create a significant wake.  Our model study thus provides strong motivation for the discovery potential of performing the analyses we have described with experimental $Z$+jet measurements.

\acknowledgments
We are grateful to Dani Pablos for providing us with hybrid model samples and much helpful advice and to  Yen-Jie Lee, Andrew Lin, Guilherme Milhano, James Mulligan, and Jesse Thaler for helpful conversations.
This work was supported by the U.S. Department of Energy, Office of Science, Office of Nuclear Physics grant DE-SC0011090.

\bibliographystyle{JHEP}
\bibliography{Zjets.bib}

\providecommand{\href}[2]{#2}\begingroup\raggedright\begin{thebibliography}{10}

\bibitem{STAR:2005gfr}
{\scshape STAR} collaboration, \emph{{Experimental and theoretical challenges
  in the search for the quark gluon plasma: The STAR Collaboration's critical
  assessment of the evidence from RHIC collisions}},
  \href{https://doi.org/10.1016/j.nuclphysa.2005.03.085}{\emph{Nucl. Phys. A}
  {\bfseries 757} (2005) 102}
  [\href{https://arxiv.org/abs/nucl-ex/0501009}{{\ttfamily nucl-ex/0501009}}].

\bibitem{PHENIX:2004vcz}
{\scshape PHENIX} collaboration, \emph{{Formation of dense partonic matter in
  relativistic nucleus-nucleus collisions at RHIC: Experimental evaluation by
  the PHENIX collaboration}},
  \href{https://doi.org/10.1016/j.nuclphysa.2005.03.086}{\emph{Nucl. Phys. A}
  {\bfseries 757} (2005) 184}
  [\href{https://arxiv.org/abs/nucl-ex/0410003}{{\ttfamily nucl-ex/0410003}}].

\bibitem{Aad:2010bu}
{\scshape ATLAS} collaboration, \emph{{Observation of a Centrality-Dependent
  Dijet Asymmetry in Lead-Lead Collisions at $\sqrt{s_{NN}}=2.77$ TeV with the
  ATLAS Detector at the LHC}},
  \href{https://doi.org/10.1103/PhysRevLett.105.252303}{\emph{Phys. Rev. Lett.}
  {\bfseries 105} (2010) 252303}
  [\href{https://arxiv.org/abs/1011.6182}{{\ttfamily 1011.6182}}].

\bibitem{Chatrchyan:2011sx}
{\scshape CMS} collaboration, \emph{{Observation and studies of jet quenching
  in PbPb collisions at nucleon-nucleon center-of-mass energy = 2.76 TeV}},
  \href{https://doi.org/10.1103/PhysRevC.84.024906}{\emph{Phys. Rev.}
  {\bfseries C84} (2011) 024906}
  [\href{https://arxiv.org/abs/1102.1957}{{\ttfamily 1102.1957}}].

\bibitem{Adam:2015ewa}
{\scshape ALICE} collaboration, \emph{{Measurement of jet suppression in
  central Pb-Pb collisions at $\sqrt{s_{\rm NN}}$ = 2.76 TeV}},
  \href{https://doi.org/10.1016/j.physletb.2015.04.039}{\emph{Phys. Lett.}
  {\bfseries B746} (2015) 1}
  [\href{https://arxiv.org/abs/1502.01689}{{\ttfamily 1502.01689}}].

\bibitem{Casalderrey-Solana:2007knd}
J.~Casalderrey-Solana and C.A.~Salgado, \emph{{Introductory lectures on jet
  quenching in heavy ion collisions}}, {\emph{Acta Phys. Polon. B} {\bfseries
  38} (2007) 3731} [\href{https://arxiv.org/abs/0712.3443}{{\ttfamily
  0712.3443}}].

\bibitem{dEnterria:2009xfs}
D.~d'Enterria, \emph{{Jet quenching}},
  \href{https://doi.org/10.1007/978-3-642-01539-7_16}{\emph{Landolt-Bornstein}
  {\bfseries 23} (2010) 471} [\href{https://arxiv.org/abs/0902.2011}{{\ttfamily
  0902.2011}}].

\bibitem{Wiedemann:2009sh}
U.A.~Wiedemann, \emph{{Jet Quenching in Heavy Ion Collisions}},
  \href{https://doi.org/10.1007/978-3-642-01539-7_17}{\emph{Landolt-Bornstein}
  {\bfseries 23} (2010) 521} [\href{https://arxiv.org/abs/0908.2306}{{\ttfamily
  0908.2306}}].

\bibitem{Majumder:2010qh}
A.~Majumder and M.~Van~Leeuwen, \emph{{The Theory and Phenomenology of
  Perturbative QCD Based Jet Quenching}},
  \href{https://doi.org/10.1016/j.ppnp.2010.09.001}{\emph{Prog. Part. Nucl.
  Phys.} {\bfseries 66} (2011) 41}
  [\href{https://arxiv.org/abs/1002.2206}{{\ttfamily 1002.2206}}].

\bibitem{Mehtar-Tani:2013pia}
Y.~Mehtar-Tani, J.G.~Milhano and K.~Tywoniuk, \emph{{Jet physics in heavy-ion
  collisions}}, \href{https://doi.org/10.1142/S0217751X13400137}{\emph{Int. J.
  Mod. Phys.} {\bfseries A28} (2013) 1340013}
  [\href{https://arxiv.org/abs/1302.2579}{{\ttfamily 1302.2579}}].

\bibitem{Connors:2017ptx}
M.~Connors, C.~Nattrass, R.~Reed and S.~Salur, \emph{{Jet measurements in heavy
  ion physics}}, \href{https://doi.org/10.1103/RevModPhys.90.025005}{\emph{Rev.
  Mod. Phys.} {\bfseries 90} (2018) 025005}
  [\href{https://arxiv.org/abs/1705.01974}{{\ttfamily 1705.01974}}].

\bibitem{Busza:2018rrf}
W.~Busza, K.~Rajagopal and W.~van~der Schee, \emph{{Heavy Ion Collisions: The
  Big Picture, and the Big Questions}},
  \href{https://doi.org/10.1146/annurev-nucl-101917-020852}{\emph{Ann. Rev.
  Nucl. Part. Sci.} {\bfseries 68} (2018) 339}
  [\href{https://arxiv.org/abs/1802.04801}{{\ttfamily 1802.04801}}].

\bibitem{Baier:1996kr}
R.~Baier, Y.L.~Dokshitzer, A.H.~Mueller, S.~Peigne and D.~Schiff,
  \emph{{Radiative energy loss of high-energy quarks and gluons in a finite
  volume quark - gluon plasma}},
  \href{https://doi.org/10.1016/S0550-3213(96)00553-6}{\emph{Nucl. Phys. B}
  {\bfseries 483} (1997) 291}
  [\href{https://arxiv.org/abs/hep-ph/9607355}{{\ttfamily hep-ph/9607355}}].

\bibitem{Zakharov:1997uu}
B.G.~Zakharov, \emph{{Radiative energy loss of high-energy quarks in finite
  size nuclear matter and quark - gluon plasma}},
  \href{https://doi.org/10.1134/1.567389}{\emph{JETP Lett.} {\bfseries 65}
  (1997) 615} [\href{https://arxiv.org/abs/hep-ph/9704255}{{\ttfamily
  hep-ph/9704255}}].

\bibitem{Gyulassy:2000er}
M.~Gyulassy, P.~Levai and I.~Vitev, \emph{{Reaction operator approach to
  nonAbelian energy loss}},
  \href{https://doi.org/10.1016/S0550-3213(00)00652-0}{\emph{Nucl. Phys. B}
  {\bfseries 594} (2001) 371}
  [\href{https://arxiv.org/abs/nucl-th/0006010}{{\ttfamily nucl-th/0006010}}].

\bibitem{Armesto:2003jh}
N.~Armesto, C.A.~Salgado and U.A.~Wiedemann, \emph{{Medium induced gluon
  radiation off massive quarks fills the dead cone}},
  \href{https://doi.org/10.1103/PhysRevD.69.114003}{\emph{Phys. Rev. D}
  {\bfseries 69} (2004) 114003}
  [\href{https://arxiv.org/abs/hep-ph/0312106}{{\ttfamily hep-ph/0312106}}].

\bibitem{Gubser:2006bz}
S.S.~Gubser, \emph{{Drag force in AdS/CFT}},
  \href{https://doi.org/10.1103/PhysRevD.74.126005}{\emph{Phys. Rev. D}
  {\bfseries 74} (2006) 126005}
  [\href{https://arxiv.org/abs/hep-th/0605182}{{\ttfamily hep-th/0605182}}].

\bibitem{Casalderrey-Solana:2006fio}
J.~Casalderrey-Solana and D.~Teaney, \emph{{Heavy quark diffusion in strongly
  coupled N=4 Yang-Mills}},
  \href{https://doi.org/10.1103/PhysRevD.74.085012}{\emph{Phys. Rev. D}
  {\bfseries 74} (2006) 085012}
  [\href{https://arxiv.org/abs/hep-ph/0605199}{{\ttfamily hep-ph/0605199}}].

\bibitem{Herzog:2006gh}
C.P.~Herzog, A.~Karch, P.~Kovtun, C.~Kozcaz and L.G.~Yaffe, \emph{{Energy loss
  of a heavy quark moving through N=4 supersymmetric Yang-Mills plasma}},
  \href{https://doi.org/10.1088/1126-6708/2006/07/013}{\emph{JHEP} {\bfseries
  07} (2006) 013} [\href{https://arxiv.org/abs/hep-th/0605158}{{\ttfamily
  hep-th/0605158}}].

\bibitem{Liu:2006ug}
H.~Liu, K.~Rajagopal and U.A.~Wiedemann, \emph{{Calculating the jet quenching
  parameter from AdS/CFT}},
  \href{https://doi.org/10.1103/PhysRevLett.97.182301}{\emph{Phys. Rev. Lett.}
  {\bfseries 97} (2006) 182301}
  [\href{https://arxiv.org/abs/hep-ph/0605178}{{\ttfamily hep-ph/0605178}}].

\bibitem{Chesler:2008uy}
P.M.~Chesler, K.~Jensen, A.~Karch and L.G.~Yaffe, \emph{{Light quark energy
  loss in strongly-coupled N = 4 supersymmetric Yang-Mills plasma}},
  \href{https://doi.org/10.1103/PhysRevD.79.125015}{\emph{Phys. Rev. D}
  {\bfseries 79} (2009) 125015}
  [\href{https://arxiv.org/abs/0810.1985}{{\ttfamily 0810.1985}}].

\bibitem{Casalderrey-Solana:2011dxg}
J.~Casalderrey-Solana, H.~Liu, D.~Mateos, K.~Rajagopal and U.A.~Wiedemann,
  \emph{{Gauge/String Duality, Hot QCD and Heavy Ion Collisions}}, Cambridge
  University Press (2014),
  \href{https://doi.org/10.1017/CBO9781139136747}{10.1017/CBO9781139136747},
  [\href{https://arxiv.org/abs/1101.0618}{{\ttfamily 1101.0618}}].

\bibitem{Casalderrey-Solana:2004fdk}
J.~Casalderrey-Solana, E.V.~Shuryak and D.~Teaney, \emph{{Conical flow induced
  by quenched QCD jets}},
  \href{https://doi.org/10.1088/1742-6596/27/1/003}{\emph{J. Phys. Conf. Ser.}
  {\bfseries 27} (2005) 22}
  [\href{https://arxiv.org/abs/hep-ph/0411315}{{\ttfamily hep-ph/0411315}}].

\bibitem{Neufeld:2008fi}
R.B.~Neufeld, B.~Muller and J.~Ruppert, \emph{{Sonic Mach Cones Induced by Fast
  Partons in a Perturbative Quark-Gluon Plasma}},
  \href{https://doi.org/10.1103/PhysRevC.78.041901}{\emph{Phys. Rev. C}
  {\bfseries 78} (2008) 041901}
  [\href{https://arxiv.org/abs/0802.2254}{{\ttfamily 0802.2254}}].

\bibitem{Betz:2008ka}
B.~Betz, J.~Noronha, G.~Torrieri, M.~Gyulassy, I.~Mishustin and D.H.~Rischke,
  \emph{{Universality of the Diffusion Wake from Stopped and Punch-Through Jets
  in Heavy-Ion Collisions}},
  \href{https://doi.org/10.1103/PhysRevC.79.034902}{\emph{Phys. Rev. C}
  {\bfseries 79} (2009) 034902}
  [\href{https://arxiv.org/abs/0812.4401}{{\ttfamily 0812.4401}}].

\bibitem{Casalderrey-Solana:2016jvj}
J.~Casalderrey-Solana, D.~Gulhan, G.~Milhano, D.~Pablos and K.~Rajagopal,
  \emph{{Angular Structure of Jet Quenching Within a Hybrid Strong/Weak
  Coupling Model}}, \href{https://doi.org/10.1007/JHEP03(2017)135}{\emph{JHEP}
  {\bfseries 03} (2017) 135}
  [\href{https://arxiv.org/abs/1609.05842}{{\ttfamily 1609.05842}}].

\bibitem{Tachibana:2017syd}
Y.~Tachibana, N.-B.~Chang and G.-Y.~Qin, \emph{{Full jet in quark-gluon plasma
  with hydrodynamic medium response}},
  \href{https://doi.org/10.1103/PhysRevC.95.044909}{\emph{Phys. Rev. C}
  {\bfseries 95} (2017) 044909}
  [\href{https://arxiv.org/abs/1701.07951}{{\ttfamily 1701.07951}}].

\bibitem{JETSCAPE:2020uew}
{\scshape JETSCAPE} collaboration, \emph{{Hydrodynamic response to jets with a
  source based on causal diffusion}},
  \href{https://doi.org/10.1016/j.nuclphysa.2020.121920}{\emph{Nucl. Phys. A}
  {\bfseries 1005} (2021) 121920}
  [\href{https://arxiv.org/abs/2002.12250}{{\ttfamily 2002.12250}}].

\bibitem{Casalderrey-Solana:2020rsj}
J.~Casalderrey-Solana, J.G.~Milhano, D.~Pablos, K.~Rajagopal and X.~Yao,
  \emph{{Jet Wake from Linearized Hydrodynamics}},
  \href{https://doi.org/10.1007/JHEP05(2021)230}{\emph{JHEP} {\bfseries 05}
  (2021) 230} [\href{https://arxiv.org/abs/2010.01140}{{\ttfamily
  2010.01140}}].

\bibitem{Dreyer:2018nbf}
F.A.~Dreyer, G.P.~Salam and G.~Soyez, \emph{{The Lund Jet Plane}},
  \href{https://doi.org/10.1007/JHEP12(2018)064}{\emph{JHEP} {\bfseries 12}
  (2018) 064} [\href{https://arxiv.org/abs/1807.04758}{{\ttfamily
  1807.04758}}].

\bibitem{Andrews:2018jcm}
H.A.~Andrews et~al., \emph{{Novel tools and observables for jet physics in
  heavy-ion collisions}},
  \href{https://doi.org/10.1088/1361-6471/ab7cbc}{\emph{J. Phys. G} {\bfseries
  47} (2020) 065102} [\href{https://arxiv.org/abs/1808.03689}{{\ttfamily
  1808.03689}}].

\bibitem{Mehtar-Tani:2016aco}
Y.~Mehtar-Tani and K.~Tywoniuk, \emph{{Groomed jets in heavy-ion collisions:
  sensitivity to medium-induced bremsstrahlung}},
  \href{https://doi.org/10.1007/JHEP04(2017)125}{\emph{JHEP} {\bfseries 04}
  (2017) 125} [\href{https://arxiv.org/abs/1610.08930}{{\ttfamily
  1610.08930}}].

\bibitem{Casalderrey-Solana:2019ubu}
J.~Casalderrey-Solana, G.~Milhano, D.~Pablos and K.~Rajagopal,
  \emph{{Modification of Jet Substructure in Heavy Ion Collisions as a Probe of
  the Resolution Length of Quark-Gluon Plasma}},
  \href{https://doi.org/10.1007/JHEP01(2020)044}{\emph{JHEP} {\bfseries 01}
  (2020) 044} [\href{https://arxiv.org/abs/1907.11248}{{\ttfamily
  1907.11248}}].

\bibitem{Mehtar-Tani:2019rrk}
Y.~Mehtar-Tani, A.~Soto-Ontoso and K.~Tywoniuk, \emph{{Dynamical grooming of
  QCD jets}}, \href{https://doi.org/10.1103/PhysRevD.101.034004}{\emph{Phys.
  Rev. D} {\bfseries 101} (2020) 034004}
  [\href{https://arxiv.org/abs/1911.00375}{{\ttfamily 1911.00375}}].

\bibitem{Apolinario:2020uvt}
L.~Apolin\'ario, A.~Cordeiro and K.~Zapp, \emph{{Time reclustering for jet
  quenching studies}},
  \href{https://doi.org/10.1140/epjc/s10052-021-09346-8}{\emph{Eur. Phys. J. C}
  {\bfseries 81} (2021) 561}
  [\href{https://arxiv.org/abs/2012.02199}{{\ttfamily 2012.02199}}].

\bibitem{Milhano:2017nzm}
G.~Milhano, U.A.~Wiedemann and K.C.~Zapp, \emph{{Sensitivity of jet
  substructure to jet-induced medium response}},
  \href{https://doi.org/10.1016/j.physletb.2018.01.029}{\emph{Phys. Lett. B}
  {\bfseries 779} (2018) 409}
  [\href{https://arxiv.org/abs/1707.04142}{{\ttfamily 1707.04142}}].

\bibitem{CMS:2021vui}
{\scshape CMS} collaboration, \emph{{First measurement of large area jet
  transverse momentum spectra in heavy-ion collisions}},
  \href{https://doi.org/10.1007/JHEP05(2021)284}{\emph{JHEP} {\bfseries 05}
  (2021) 284} [\href{https://arxiv.org/abs/2102.13080}{{\ttfamily
  2102.13080}}].

\bibitem{Larkoski:2014wba}
A.J.~Larkoski, S.~Marzani, G.~Soyez and J.~Thaler, \emph{{Soft Drop}},
  \href{https://doi.org/10.1007/JHEP05(2014)146}{\emph{JHEP} {\bfseries 05}
  (2014) 146} [\href{https://arxiv.org/abs/1402.2657}{{\ttfamily 1402.2657}}].

\bibitem{Dasgupta:2013ihk}
M.~Dasgupta, A.~Fregoso, S.~Marzani and G.P.~Salam, \emph{{Towards an
  understanding of jet substructure}},
  \href{https://doi.org/10.1007/JHEP09(2013)029}{\emph{JHEP} {\bfseries 09}
  (2013) 029} [\href{https://arxiv.org/abs/1307.0007}{{\ttfamily 1307.0007}}].

\bibitem{Larkoski:2015lea}
A.J.~Larkoski, S.~Marzani and J.~Thaler, \emph{{Sudakov Safety in Perturbative
  QCD}}, \href{https://doi.org/10.1103/PhysRevD.91.111501}{\emph{Phys. Rev. D}
  {\bfseries 91} (2015) 111501}
  [\href{https://arxiv.org/abs/1502.01719}{{\ttfamily 1502.01719}}].

\bibitem{ATLAS:2019mgf}
{\scshape ATLAS} collaboration, \emph{{Measurement of soft-drop jet observables
  in $pp$ collisions with the ATLAS detector at $\sqrt {s}$ =13 TeV}},
  \href{https://doi.org/10.1103/PhysRevD.101.052007}{\emph{Phys. Rev. D}
  {\bfseries 101} (2020) 052007}
  [\href{https://arxiv.org/abs/1912.09837}{{\ttfamily 1912.09837}}].

\bibitem{ATLAS:2018zhf}
{\scshape ATLAS} collaboration, \emph{{Properties of $g\rightarrow b\bar{b}$ at
  small opening angles in $pp$ collisions with the ATLAS detector at
  $\sqrt{s}=13$ TeV}},
  \href{https://doi.org/10.1103/PhysRevD.99.052004}{\emph{Phys. Rev. D}
  {\bfseries 99} (2019) 052004}
  [\href{https://arxiv.org/abs/1812.09283}{{\ttfamily 1812.09283}}].

\bibitem{CMS:2018ypj}
{\scshape CMS} collaboration, \emph{{Measurement of jet substructure
  observables in $\mathrm{t\overline{t}}$ events from proton-proton collisions
  at $\sqrt{s}=$ 13TeV}},
  \href{https://doi.org/10.1103/PhysRevD.98.092014}{\emph{Phys. Rev. D}
  {\bfseries 98} (2018) 092014}
  [\href{https://arxiv.org/abs/1808.07340}{{\ttfamily 1808.07340}}].

\bibitem{Mulligan:2020tim}
J.~Mulligan and M.~Ploskon, \emph{{Identifying groomed jet splittings in
  heavy-ion collisions}},
  \href{https://doi.org/10.1103/PhysRevC.102.044913}{\emph{Phys. Rev. C}
  {\bfseries 102} (2020) 044913}
  [\href{https://arxiv.org/abs/2006.01812}{{\ttfamily 2006.01812}}].

\bibitem{Mehtar-Tani:2010ebp}
Y.~Mehtar-Tani, C.A.~Salgado and K.~Tywoniuk, \emph{{Anti-angular ordering of
  gluon radiation in QCD media}},
  \href{https://doi.org/10.1103/PhysRevLett.106.122002}{\emph{Phys. Rev. Lett.}
  {\bfseries 106} (2011) 122002}
  [\href{https://arxiv.org/abs/1009.2965}{{\ttfamily 1009.2965}}].

\bibitem{Chien:2016led}
Y.-T.~Chien and I.~Vitev, \emph{{Probing the Hardest Branching within Jets in
  Heavy-Ion Collisions}},
  \href{https://doi.org/10.1103/PhysRevLett.119.112301}{\emph{Phys. Rev. Lett.}
  {\bfseries 119} (2017) 112301}
  [\href{https://arxiv.org/abs/1608.07283}{{\ttfamily 1608.07283}}].

\bibitem{Chang:2017gkt}
N.-B.~Chang, S.~Cao and G.-Y.~Qin, \emph{{Probing medium-induced jet splitting
  and energy loss in heavy-ion collisions}},
  \href{https://doi.org/10.1016/j.physletb.2018.04.019}{\emph{Phys. Lett. B}
  {\bfseries 781} (2018) 423}
  [\href{https://arxiv.org/abs/1707.03767}{{\ttfamily 1707.03767}}].

\bibitem{Chien:2018dfn}
Y.-T.~Chien and R.~Kunnawalkam~Elayavalli, \emph{{Probing heavy ion collisions
  using quark and gluon jet substructure}},
  \href{https://arxiv.org/abs/1803.03589}{{\ttfamily 1803.03589}}.

\bibitem{Caucal:2019uvr}
P.~Caucal, E.~Iancu and G.~Soyez, \emph{{Deciphering the $z_g$ distribution in
  ultrarelativistic heavy ion collisions}},
  \href{https://doi.org/10.1007/JHEP10(2019)273}{\emph{JHEP} {\bfseries 10}
  (2019) 273} [\href{https://arxiv.org/abs/1907.04866}{{\ttfamily
  1907.04866}}].

\bibitem{CMS:2017qlm}
{\scshape CMS} collaboration, \emph{{Measurement of the Splitting Function in
  $pp$ and Pb-Pb Collisions at $\sqrt{s_{_{\mathrm{NN}}}} =$ 5.02 TeV}},
  \href{https://doi.org/10.1103/PhysRevLett.120.142302}{\emph{Phys. Rev. Lett.}
  {\bfseries 120} (2018) 142302}
  [\href{https://arxiv.org/abs/1708.09429}{{\ttfamily 1708.09429}}].

\bibitem{CMS:2018fof}
{\scshape CMS} collaboration, \emph{{Measurement of the groomed jet mass in
  PbPb and pp collisions at $ \sqrt{s_{\mathrm{NN}}}=5.02 $ TeV}},
  \href{https://doi.org/10.1007/JHEP10(2018)161}{\emph{JHEP} {\bfseries 10}
  (2018) 161} [\href{https://arxiv.org/abs/1805.05145}{{\ttfamily
  1805.05145}}].

\bibitem{ALICE:2021obz}
{\scshape ALICE} collaboration, \emph{{Measurement of the groomed jet radius
  and momentum splitting fraction in pp and Pb$-$Pb collisions at $\sqrt{s_{\rm
  NN}} = 5.02$ TeV}},  \href{https://arxiv.org/abs/2107.12984}{{\ttfamily
  2107.12984}}.

\bibitem{ALICE:2019ykw}
{\scshape ALICE} collaboration, \emph{{Exploration of jet substructure using
  iterative declustering in pp and Pb\textendash{}Pb collisions at LHC
  energies}}, \href{https://doi.org/10.1016/j.physletb.2020.135227}{\emph{Phys.
  Lett. B} {\bfseries 802} (2020) 135227}
  [\href{https://arxiv.org/abs/1905.02512}{{\ttfamily 1905.02512}}].

\bibitem{Oh:2020yyn}
{\scshape STAR} collaboration, \emph{{Jet shapes and fragmentation functions in
  Au+Au collisions at $\sqrt {s_{NN}}$ = 200GeV in STAR}},
  \href{https://doi.org/10.1016/j.nuclphysa.2020.121808}{\emph{Nucl. Phys. A}
  {\bfseries 1005} (2021) 121808}
  [\href{https://arxiv.org/abs/2002.06217}{{\ttfamily 2002.06217}}].

\bibitem{Renk:2012ve}
T.~Renk, \emph{{Biased showers: A common conceptual framework for the
  interpretation of high-$P_T$ observables in heavy-ion collisions}},
  \href{https://doi.org/10.1103/PhysRevC.88.054902}{\emph{Phys. Rev. C}
  {\bfseries 88} (2013) 054902}
  [\href{https://arxiv.org/abs/1212.0646}{{\ttfamily 1212.0646}}].

\bibitem{Milhano:2015mng}
J.G.~Milhano and K.C.~Zapp, \emph{{Origins of the di-jet asymmetry in heavy ion
  collisions}},
  \href{https://doi.org/10.1140/epjc/s10052-016-4130-9}{\emph{Eur. Phys. J.}
  {\bfseries C76} (2016) 288}
  [\href{https://arxiv.org/abs/1512.08107}{{\ttfamily 1512.08107}}].

\bibitem{Spousta:2015fca}
M.~Spousta and B.~Cole, \emph{{Interpreting single jet measurements in Pb $+$
  Pb collisions at the LHC}},
  \href{https://doi.org/10.1140/epjc/s10052-016-3896-0}{\emph{Eur. Phys. J. C}
  {\bfseries 76} (2016) 50} [\href{https://arxiv.org/abs/1504.05169}{{\ttfamily
  1504.05169}}].

\bibitem{Rajagopal:2016uip}
K.~Rajagopal, A.V.~Sadofyev and W.~van~der Schee, \emph{{Evolution of the jet
  opening angle distribution in holographic plasma}},
  \href{https://doi.org/10.1103/PhysRevLett.116.211603}{\emph{Phys. Rev. Lett.}
  {\bfseries 116} (2016) 211603}
  [\href{https://arxiv.org/abs/1602.04187}{{\ttfamily 1602.04187}}].

\bibitem{Brewer:2017fqy}
J.~Brewer, K.~Rajagopal, A.~Sadofyev and W.~Van Der~Schee, \emph{{Evolution of
  the Mean Jet Shape and Dijet Asymmetry Distribution of an Ensemble of
  Holographic Jets in Strongly Coupled Plasma}},
  \href{https://doi.org/10.1007/JHEP02(2018)015}{\emph{JHEP} {\bfseries 02}
  (2018) 015} [\href{https://arxiv.org/abs/1710.03237}{{\ttfamily
  1710.03237}}].

\bibitem{Brewer:2018mpk}
J.~Brewer, A.~Sadofyev and W.~van~der Schee, \emph{{Jet shape modifications in
  holographic dijet systems}},
  \href{https://doi.org/10.1016/j.physletb.2021.136492}{\emph{Phys. Lett. B}
  {\bfseries 820} (2021) 136492}
  [\href{https://arxiv.org/abs/1809.10695}{{\ttfamily 1809.10695}}].

\bibitem{Casalderrey-Solana:2018wrw}
J.~Casalderrey-Solana, Z.~Hulcher, G.~Milhano, D.~Pablos and K.~Rajagopal,
  \emph{{Simultaneous description of hadron and jet suppression in heavy-ion
  collisions}}, \href{https://doi.org/10.1103/PhysRevC.99.051901}{\emph{Phys.
  Rev. C} {\bfseries 99} (2019) 051901}
  [\href{https://arxiv.org/abs/1808.07386}{{\ttfamily 1808.07386}}].

\bibitem{Caucal:2020xad}
P.~Caucal, E.~Iancu, A.H.~Mueller and G.~Soyez, \emph{{Nuclear modification
  factors for jet fragmentation}},
  \href{https://doi.org/10.1007/JHEP10(2020)204}{\emph{JHEP} {\bfseries 10}
  (2020) 204} [\href{https://arxiv.org/abs/2005.05852}{{\ttfamily
  2005.05852}}].

\bibitem{Brewer:2018dfs}
J.~Brewer, J.G.~Milhano and J.~Thaler, \emph{{Sorting out quenched jets}},
  \href{https://doi.org/10.1103/PhysRevLett.122.222301}{\emph{Phys. Rev. Lett.}
  {\bfseries 122} (2019) 222301}
  [\href{https://arxiv.org/abs/1812.05111}{{\ttfamily 1812.05111}}].

\bibitem{Takacs:2021bpv}
A.~Takacs and K.~Tywoniuk, \emph{{Quenching effects in the cumulative jet
  spectrum}},  \href{https://arxiv.org/abs/2103.14676}{{\ttfamily 2103.14676}}.

\bibitem{Du:2020pmp}
Y.-L.~Du, D.~Pablos and K.~Tywoniuk, \emph{{Deep learning jet modifications in
  heavy-ion collisions}},
  \href{https://doi.org/10.1007/JHEP03(2021)206}{\emph{JHEP} {\bfseries 21}
  (2020) 206} [\href{https://arxiv.org/abs/2012.07797}{{\ttfamily
  2012.07797}}].

\bibitem{Du:2021pqa}
Y.-L.~Du, D.~Pablos and K.~Tywoniuk, \emph{{Jet tomography in heavy ion
  collisions with deep learning}},
  \href{https://arxiv.org/abs/2106.11271}{{\ttfamily 2106.11271}}.

\bibitem{CMS:2017ehl}
{\scshape CMS} collaboration, \emph{{Study of jet quenching with
  isolated-photon+jet correlations in PbPb and pp collisions at
  $\sqrt{s_{_{\mathrm{NN}}}} =$ 5.02 TeV}},
  \href{https://doi.org/10.1016/j.physletb.2018.07.061}{\emph{Phys. Lett. B}
  {\bfseries 785} (2018) 14}
  [\href{https://arxiv.org/abs/1711.09738}{{\ttfamily 1711.09738}}].

\bibitem{CMS:2017eqd}
{\scshape CMS} collaboration, \emph{{Study of Jet Quenching with $Z+\text{jet}$
  Correlations in Pb-Pb and $pp$ Collisions at ${\sqrt{s}}_{NN}=5.02\text{
  }\text{ }\mathrm{TeV}$}},
  \href{https://doi.org/10.1103/PhysRevLett.119.082301}{\emph{Phys. Rev. Lett.}
  {\bfseries 119} (2017) 082301}
  [\href{https://arxiv.org/abs/1702.01060}{{\ttfamily 1702.01060}}].

\bibitem{CMS:2018jco}
{\scshape CMS} collaboration, \emph{{Jet Shapes of Isolated Photon-Tagged Jets
  in Pb-Pb and pp Collisions at $\sqrt{s_\mathrm{NN}} =$ 5.02 TeV}},
  \href{https://doi.org/10.1103/PhysRevLett.122.152001}{\emph{Phys. Rev. Lett.}
  {\bfseries 122} (2019) 152001}
  [\href{https://arxiv.org/abs/1809.08602}{{\ttfamily 1809.08602}}].

\bibitem{ATLAS:2018dgb}
{\scshape ATLAS} collaboration, \emph{{Measurement of photon\textendash{}jet
  transverse momentum correlations in 5.02 TeV Pb + Pb and $pp$ collisions with
  ATLAS}}, \href{https://doi.org/10.1016/j.physletb.2018.12.023}{\emph{Phys.
  Lett. B} {\bfseries 789} (2019) 167}
  [\href{https://arxiv.org/abs/1809.07280}{{\ttfamily 1809.07280}}].

\bibitem{ATLAS:2019dsv}
{\scshape ATLAS} collaboration, \emph{{Comparison of Fragmentation Functions
  for Jets Dominated by Light Quarks and Gluons from $pp$ and Pb+Pb Collisions
  in ATLAS}}, \href{https://doi.org/10.1103/PhysRevLett.123.042001}{\emph{Phys.
  Rev. Lett.} {\bfseries 123} (2019) 042001}
  [\href{https://arxiv.org/abs/1902.10007}{{\ttfamily 1902.10007}}].

\bibitem{Sjostrand:2007gs}
T.~Sjostrand, S.~Mrenna and P.Z.~Skands, \emph{{A Brief Introduction to PYTHIA
  8.1}}, \href{https://doi.org/10.1016/j.cpc.2008.01.036}{\emph{Comput. Phys.
  Commun.} {\bfseries 178} (2008) 852}
  [\href{https://arxiv.org/abs/0710.3820}{{\ttfamily 0710.3820}}].

\bibitem{Larkoski:2013eya}
A.J.~Larkoski, G.P.~Salam and J.~Thaler, \emph{{Energy Correlation Functions
  for Jet Substructure}},
  \href{https://doi.org/10.1007/JHEP06(2013)108}{\emph{JHEP} {\bfseries 06}
  (2013) 108} [\href{https://arxiv.org/abs/1305.0007}{{\ttfamily 1305.0007}}].

\bibitem{Casalderrey-Solana:2014bpa}
J.~Casalderrey-Solana, D.C.~Gulhan, J.G.~Milhano, D.~Pablos and K.~Rajagopal,
  \emph{{A Hybrid Strong/Weak Coupling Approach to Jet Quenching}},
  \href{https://doi.org/10.1007/JHEP09(2015)175}{\emph{JHEP} {\bfseries 10}
  (2014) 019} [\href{https://arxiv.org/abs/1405.3864}{{\ttfamily 1405.3864}}].

\bibitem{Casalderrey-Solana:2015vaa}
J.~Casalderrey-Solana, D.C.~Gulhan, J.G.~Milhano, D.~Pablos and K.~Rajagopal,
  \emph{{Predictions for Boson-Jet Observables and Fragmentation Function
  Ratios from a Hybrid Strong/Weak Coupling Model for Jet Quenching}},
  \href{https://doi.org/10.1007/JHEP03(2016)053}{\emph{JHEP} {\bfseries 03}
  (2016) 053} [\href{https://arxiv.org/abs/1508.00815}{{\ttfamily
  1508.00815}}].

\bibitem{Casalderrey-Solana:2020jbx}
J.~Casalderrey-Solana, G.~Milhano, D.~Pablos and K.~Rajagopal, \emph{{Jet
  substructure modification probes the QGP resolution length}},
  \href{https://doi.org/10.1016/j.nuclphysa.2020.121904}{\emph{Nucl. Phys. A}
  {\bfseries 1005} (2021) 121904}
  [\href{https://arxiv.org/abs/2002.09193}{{\ttfamily 2002.09193}}].

\bibitem{Cacciari:2011ma}
M.~Cacciari, G.P.~Salam and G.~Soyez, \emph{{FastJet User Manual}},
  \href{https://doi.org/10.1140/epjc/s10052-012-1896-2}{\emph{Eur. Phys. J. C}
  {\bfseries 72} (2012) 1896}
  [\href{https://arxiv.org/abs/1111.6097}{{\ttfamily 1111.6097}}].

\bibitem{Cacciari:2008gp}
M.~Cacciari, G.P.~Salam and G.~Soyez, \emph{{The anti-$k_t$ jet clustering
  algorithm}}, \href{https://doi.org/10.1088/1126-6708/2008/04/063}{\emph{JHEP}
  {\bfseries 04} (2008) 063} [\href{https://arxiv.org/abs/0802.1189}{{\ttfamily
  0802.1189}}].

\bibitem{Chesler:2014jva}
P.M.~Chesler and K.~Rajagopal, \emph{{Jet quenching in strongly coupled
  plasma}}, \href{https://doi.org/10.1103/PhysRevD.90.025033}{\emph{Phys. Rev.
  D} {\bfseries 90} (2014) 025033}
  [\href{https://arxiv.org/abs/1402.6756}{{\ttfamily 1402.6756}}].

\bibitem{Chesler:2015nqz}
P.M.~Chesler and K.~Rajagopal, \emph{{On the Evolution of Jet Energy and
  Opening Angle in Strongly Coupled Plasma}},
  \href{https://doi.org/10.1007/JHEP05(2016)098}{\emph{JHEP} {\bfseries 05}
  (2016) 098} [\href{https://arxiv.org/abs/1511.07567}{{\ttfamily
  1511.07567}}].

\bibitem{KunnawalkamElayavalli:2017hxo}
R.~Kunnawalkam~Elayavalli and K.C.~Zapp, \emph{{Medium response in JEWEL and
  its impact on jet shape observables in heavy ion collisions}},
  \href{https://doi.org/10.1007/JHEP07(2017)141}{\emph{JHEP} {\bfseries 07}
  (2017) 141} [\href{https://arxiv.org/abs/1707.01539}{{\ttfamily
  1707.01539}}].

\bibitem{Zapp:2012ak}
K.C.~Zapp, F.~Krauss and U.A.~Wiedemann, \emph{{A perturbative framework for
  jet quenching}}, \href{https://doi.org/10.1007/JHEP03(2013)080}{\emph{JHEP}
  {\bfseries 03} (2013) 080} [\href{https://arxiv.org/abs/1212.1599}{{\ttfamily
  1212.1599}}].

\bibitem{Pablos:2019ngg}
D.~Pablos, \emph{{Jet Suppression From a Small to Intermediate to Large
  Radius}}, \href{https://doi.org/10.1103/PhysRevLett.124.052301}{\emph{Phys.
  Rev. Lett.} {\bfseries 124} (2020) 052301}
  [\href{https://arxiv.org/abs/1907.12301}{{\ttfamily 1907.12301}}].

\end{thebibliography}\endgroup

\end{document}